\documentclass[5p,times]{elsarticle}

\usepackage{hyperref}
\usepackage{float}
\usepackage{caption} 
\usepackage{subcaption}
\usepackage{booktabs} 
\usepackage{amsfonts,amsmath,amsthm}
\usepackage{graphicx}
\usepackage[capitalize]{cleveref}
\usepackage{balance}
\usepackage{multirow}
\usepackage{hhline}
\usepackage{siunitx}
\usepackage{pdfpages}

\makeatletter
\def\ps@pprintTitle{%
	\let\@oddhead\@empty
	\let\@evenhead\@empty
	\let\@oddfoot\@empty
	\let\@evenfoot\@oddfoot
}
\makeatother

\begin{document}

\begin{frontmatter}

\title{An optimization framework for task allocation in the edge/hub/cloud paradigm}

\author[label1,label2]{Andreas Kouloumpris\corref{cor1}}
\ead{kouloumpris.andreas@ucy.ac.cy}
\cortext[cor1]{Corresponding author.}

\author[label2]{Georgios L. Stavrinides}

\author[label1,label2]{Maria K. Michael}

\author[label1,label2]{Theocharis Theocharides}

\affiliation[label1]{organization={Department of Electrical and Computer Engineering, University of Cyprus},
            addressline={1 Panepistimiou Avenue, Aglantzia, P.O. Box 20537}, 
            city={Nicosia},
            postcode={1678}, 
            country={Cyprus}}

\affiliation[label2]{organization={KIOS Research and Innovation Center of Excellence, University of Cyprus},
            addressline={1 Panepistimiou Avenue, Aglantzia, P.O. Box 20537}, 
            city={Nicosia},
            postcode={1678}, 
            country={Cyprus}}

\begin{abstract}
With the advent of the Internet of Things (IoT), novel critical applications have emerged that leverage the edge/hub/cloud paradigm, which diverges from the conventional edge computing perspective. A growing number of such applications require a streamlined architecture for their effective execution, often comprising a single edge device with sensing capabilities, a single hub device (e.g., a laptop or smartphone) for managing and assisting the edge device, and a more computationally capable cloud server. Typical examples include the utilization of an unmanned aerial vehicle (UAV) for critical infrastructure inspection or a wearable biomedical device (e.g., a smartwatch) for remote patient monitoring. Task allocation in this streamlined architecture is particularly challenging, due to the computational, communication, and energy limitations of the devices at the network edge. Consequently, there is a need for a comprehensive framework that can address the specific task allocation problem optimally and efficiently. To this end, we propose a complete, binary integer linear programming (BILP) based formulation for an application-driven design-time approach, capable of providing an optimal task allocation in the targeted edge/hub/cloud environment. The proposed method minimizes the desired objective, either the overall latency or overall energy consumption, while considering several crucial parameters and constraints often overlooked in related literature. We evaluate our framework using a real-world use-case scenario, as well as appropriate synthetic benchmarks. Our extensive experimentation reveals that the proposed approach yields optimal and scalable results, enabling efficient design space exploration for different applications and computational devices.
\end{abstract}


\end{frontmatter}

\sloppy

\section{Introduction}
The rapid expansion of the Internet of Things (IoT) has led to the emergence of applications that leverage diverse system architectures in the edge-cloud continuum. These systems vary with respect to where the computation takes place, based on the unique requirements of each application. One such architecture follows the edge/hub/cloud paradigm, which facilitates remote decision-making and real-time response \cite{Noghabi2018, Sidhardhan2021}.
This paradigm differs from the traditional edge computing concept, which typically considers a bottom layer of sensing devices (IoT or edge devices layer), an intermediate layer with servers of moderate computational capacity (edge or fog servers layer), and a top layer with more computationally capable cloud servers (cloud layer) \cite{Iorga2018, Kekki2018}. 
In contrast, an edge/hub/cloud environment encompasses edge and hub devices in the bottom layer and cloud servers in the top layer. A hub device, such as a laptop or smartphone, interacts with the edge devices and facilitates their communication with the cloud servers. It is typically closer to the edge devices than an edge/fog server, although it is less capable. On the other hand, it generally has a higher computational capacity than an edge device. In this environment, the edge and hub devices may be battery-operated.

In the edge/hub/cloud paradigm, an application typically comprises tasks with precedence relationships among them, forming a task flow graph. An increasing number of these applications employ a streamlined architecture, consisting of an edge device, a hub device, and a cloud server.
One notable category involves the monitoring, inspection, data collection, processing, and intelligent real-time decision-making in critical infrastructures and environments, utilizing a single unmanned aerial vehicle (UAV). 
In addition to the UAV's sensing abilities, its embedded computational and communication capabilities allow it to act as an edge device, typically communicating with a ground station (i.e., a hub device) that also has computational capabilities. The hub device (e.g., a laptop) can in turn communicate with a cloud server that can run the more computationally intensive tasks \cite{Cheng2021, icarusTheocharides}. 
Another class of applications that leverage this streamlined edge/hub/cloud architecture is in the tele-healthcare domain, where a wearable biomedical edge device (e.g., a smartwatch) can assist in the remote monitoring of patients. Such a device can continuously track a patient's vital signs for the detection of alarming indicators, while interacting with a hub device (e.g., a smartphone), which in turn can communicate with a cloud server for further data analysis \cite{Dao2023}.
In these use-cases, a single edge device, a single hub device, and a single cloud server suffice for handling the processing and communication requirements of the application.

Even though the considered edge/hub/cloud architecture is streamlined, it involves many challenges. Moving from the cloud towards the edge of the network, the computational and communication capabilities, as well as the memory, storage, and energy capacities of the devices, become more limited. 
Hence, optimal task allocation, necessitated by the critical nature of the examined applications, can become particularly challenging. This challenge is intensified in operating environments with diverse communication characteristics, where such applications are often deployed. Moreover, these applications tend to incorporate computationally demanding algorithms, like machine learning inference and time-bound stream processing techniques, which further complicate the allocation of their tasks.
To our knowledge, the task allocation approaches proposed so far in the literature do not consider this particular architecture, nor a complete set of parameters that typically characterize such environments. 

Consequently, we propose a novel formulation, integrated in an application-driven design-time (i.e., offline) framework, to solve the task allocation problem in the examined edge/hub/cloud architecture optimally and efficiently. We consider several crucial parameters and constraints that are usually ignored in related research efforts, such as the computational and communication latency and energy required for the processing of the tasks, as well as the memory, storage, and energy limitations of the devices. The proposed approach supports two distinct optimization objectives, the minimization of either latency or energy, enabling design space exploration with respect to different devices and their connectivity.
The main contributions of this work, which builds upon our preliminary research in \cite{Kouloumpris2019}, are summarized as follows:
\begin{enumerate}
        \item We present a comprehensive binary integer linear programming (BILP) based formulation that encapsulates a complete set of parameters and constraints characterizing an edge/hub/cloud environment. It aims to allocate the application tasks across a streamlined architecture, while minimizing the overall latency or energy consumption.
    
        \item We propose a transformation of a task flow graph into an extended task flow graph, encompassing both the application and system models, as well as the employed energy model, to facilitate the problem formulation.  

        \item We evaluate our framework utilizing a real-world use-case scenario for the UAV-based aerial inspection of power transmission towers and lines.
        
        \item To further validate our approach and investigate its scalability to different applications, we use appropriate synthetic benchmarks with suitable parameters, which we developed and transformed for this purpose.
\end{enumerate}

The rest of the paper is arranged as follows. \cref{background} provides an overview of related research. \cref{framework} describes the proposed optimization framework. \cref{evaluation} presents the experimental setup and the evaluation results. Finally, \cref{conc} summarizes and concludes the paper.

\section{Related work}\label{background} 
Task allocation is a well-studied problem, posing ongoing challenges in various computing environments \cite{Stavrinides2019, Genez2020, Jayanetti2022, Kanbar2022, Kritikakou2022, Peixoto2022, Mo2023}.
However, previous related research efforts do not consider the edge/hub/cloud architecture, nor all of the parameters investigated in this work. This is demonstrated in \cref{table:comparison}, which summarizes our qualitative comparison with relevant state-of-the-art approaches. 
The comparison is made with respect to the objectives and parameters considered in this work, the applicability of each approach to applications comprising multiple tasks with precedence relationships among them (i.e., applications with a task flow graph structure), as well as the optimality of the solution provided by each method.
An overview of the related literature, as well as a comparison with our preliminary research, are provided in the remainder of this section.

\begin{table*}[!ht]
\centering
\caption{Qualitative comparison of this work with relevant research efforts.}
\label{table:comparison}
\footnotesize
\resizebox{0.85\textwidth}{!}{
    \begin{tabular}{@{\extracolsep{4pt}}lcccccccccc@{}} 
        \toprule
        \multirow{3}{*}{Reference} & \multicolumn{2}{c}{Objectives} & \multicolumn{6}{c}{Considered Parameters} & \multirow{2}{*}{Task Flow} & \multirow{2}{*}{Optimal}\\
         \cline{2-3}   \cline{4-9} 
        & Latency & Energy & Comp. & Comp. & Comm. &  Comm. &  \multirow{2}{*}{Memory} & \multirow{2}{*}{Storage} & \multirow{2}{*}{Graph} & \multirow{2}{*}{Solution}\\
        & Min. & Min. & Latency & Energy & Latency & Energy & & & & \\
        
        \hline
        \cite{Alfakih2021}              & \checkmark    & -           & \checkmark  & -           & -          & -          & \checkmark  & \checkmark & -          & -           \\
        \cite{Guevara2022}              & \checkmark    & -           & \checkmark  & -           & \checkmark & -          & \checkmark  & \checkmark & \checkmark & -           \\
        \cite{Weikert2022}              & \checkmark    & -           & \checkmark  & -           & \checkmark & \checkmark & \checkmark  & -          & \checkmark & -           \\
        \cite{Lai2022}                  & \checkmark    & -           & \checkmark  & -           & \checkmark & -          & \checkmark  & \checkmark & -          & -           \\
        \cite{Barijough2019}            & \checkmark    & -           & \checkmark   & -          & \checkmark & \checkmark & -           & -          & \checkmark & \checkmark  \\
        \cite{Tang2022}                 & \checkmark    & -           & \checkmark   & -          & \checkmark & -          & -           & \checkmark & -          & \checkmark  \\
        \cite{Kuang2021}                & \checkmark    & -           & \checkmark   & \checkmark & \checkmark & \checkmark & -           & -          & -          & -           \\
        
        \cite{Avgeris2022}              & -             & \checkmark  & \checkmark  & \checkmark  &\checkmark & -           & -           & -          & -          & \checkmark \\
        \cite{Khalil2018}               & -             & \checkmark  & -           & \checkmark  & -         & \checkmark  & -           & -          & -          & -          \\
        \cite{Kritikakou2023}           & -             & \checkmark  & \checkmark  & \checkmark  & -         & -           & -           & -          & \checkmark & -          \\
        \cite{Hu2020}                   & -             & \checkmark & \checkmark   & \checkmark & \checkmark & \checkmark & -          & -            & -          & -          \\
        \cite{Azizi2022}                & -             & \checkmark & \checkmark   & \checkmark & \checkmark & -          & -          & -            & -          & -          \\
        \cite{Li2022}                   & -             & \checkmark & \checkmark   & \checkmark & \checkmark & \checkmark & -          & -            & -          & -          \\

        \cite{Zhang2021}                & \checkmark    & \checkmark & \checkmark   & \checkmark & \checkmark & \checkmark & -          & -            & -          & -          \\
        \cite{Dinh2017}                 & \checkmark    & \checkmark & \checkmark   & \checkmark & \checkmark & \checkmark & -          & -            & -          & -          \\        
        \cite{Tong2023}                 & \checkmark    & \checkmark & \checkmark   & \checkmark & \checkmark & \checkmark & -          & -            & -          & -          \\

        This work & \checkmark & \checkmark & \checkmark & \checkmark & \checkmark & \checkmark & \checkmark & \checkmark & \checkmark  & \checkmark \\
        \bottomrule
    \end{tabular}
}
\end{table*}

\subsection{Latency minimization}
A number of works on task allocation in edge computing and multi-tier environments have a primary focus on latency minimization.
For instance, Alfakih et al. \cite{Alfakih2021} explore the minimization of the computational latency of task execution in an edge computing system, based on an accelerated particle swarm optimization algorithm combined with a dynamic programming approach.
Guevara et al. \cite{Guevara2022} present a reinforcement learning-based resource allocation technique for minimizing the total execution time of tasks in a fog-cloud environment.
On the other hand, Weikert et al. \cite{Weikert2022} propose an algorithm for task allocation in an IoT platform, aiming to optimize the overall latency.
Furthermore, Lai et al. \cite{Lai2022} propose an online Lyapunov optimization-based method to tackle the problem of allocating user tasks in an edge computing environment, utilizing a stochastic approach. 
Barijough et al. \cite{Barijough2019} introduce a technique for allocating real-time streaming applications under latency and quality constraints.  
Tang et al. \cite{Tang2022} propose a framework for managing the physical resources of the edge and cloud layers, so that the response time is minimized and the system throughput is improved.
Moreover, Kuang et al. \cite{Kuang2021} present an iterative algorithm based on Lagrangian dual decomposition in order to minimize latency in an edge computing system.

\subsection{Energy consumption minimization}
Several studies are focused on task allocation strategies aiming to reduce the total energy consumption.
Specifically, Avgeris et al. \cite{Avgeris2022} propose a resource allocation technique based on mixed integer linear programming in order to minimize the energy consumption of edge servers.
Within this context, Khalil et al. \cite{Khalil2018} present a framework for energy-efficient task allocation in an IoT environment, utilizing evolutionary-based meta-heuristics. 
Cui et al. \cite{Kritikakou2023} propose a heuristic algorithm for minimizing the total energy consumption of a platform comprising homogeneous processors, utilizing dynamic voltage and frequency scaling (DVFS).  
On the other hand, Hu et al. \cite{Hu2020} introduce a game-theoretic approach for task allocation in an edge computing environment to minimize the system energy consumption within an acceptable delay range.
Similarly, Azizi et al. \cite{Azizi2022} propose two priority-aware semi-greedy algorithms for allocating  IoT tasks in a heterogeneous fog platform, so that the total energy consumption is optimized, while meeting the deadline of each task.
Furthermore, Li et al. \cite{Li2022} examine a two-stage iterative algorithm, in which the resource allocation problem is decomposed into two sub-problems to obtain a suboptimal solution.

\subsection{Latency and energy consumption minimization}
On the other hand, certain related works consider both optimization objectives, the minimization of latency and energy consumption.
For instance, Zhang et al. \cite{Zhang2021} present a game theory-based scheme for task allocation in a UAV-assisted edge computing environment. The goal of the proposed approach is to minimize the weighted latency and energy consumption of the system, considering resource allocation constraints.
Dinh et al. \cite{Dinh2017} propose a semi-definite relaxation-based optimization framework for allocating tasks in an edge architecture. The particular framework aims to minimize the total latency of the tasks, as well as the total energy consumption of the system.
On the other hand, Tong et al. \cite{Tong2023} present a latency and energy-aware Stackelberg game-based task allocation strategy, considering an edge device with limited computational resources.

\subsection{Our approach vs. state-of-the-art}
Overall, none of the aforementioned research efforts considers the specific edge/hub/cloud architecture examined in this work. 
Furthermore, some approaches do not take into account the energy required for the execution of the tasks \cite{Alfakih2021, Guevara2022, Weikert2022, Lai2022, Barijough2019, Tang2022} or the energy consumed for inter-task communication \cite{Alfakih2021, Guevara2022, Lai2022, Tang2022, Avgeris2022, Kritikakou2023, Azizi2022}. 
The majority of the related studies consider devices with unlimited resources, such as memory \cite{Barijough2019, Tang2022, Kuang2021, Avgeris2022, Khalil2018, Kritikakou2023, Hu2020, Azizi2022, Li2022} and storage \cite{Weikert2022, Barijough2019, Kuang2021, Avgeris2022, Khalil2018, Kritikakou2023, Hu2020, Azizi2022, Li2022}, an assumption that is not realistic, especially in the case of resource-limited devices at the edge of the network.     
Moreover, several approaches are only applicable to single-task applications \cite{Alfakih2021, Lai2022, Tang2022, Kuang2021, Avgeris2022, Khalil2018, Hu2020, Azizi2022, Li2022} or cannot provide an optimal solution to each of the objectives considered in this work \cite{Alfakih2021, Guevara2022, Weikert2022, Lai2022, Kuang2021, Khalil2018, Kritikakou2023, Hu2020, Azizi2022, Li2022}.

Related studies that are closer to ours \cite{Zhang2021, Dinh2017, Tong2023}, even though they consider both the latency and energy aspects of the problem, do not take into account the memory and storage limitations of the devices. Furthermore, they cannot be applied to applications with precedence relationships among their tasks, and can only provide suboptimal solutions.
Hence, our proposed approach aims to fill these gaps, by incorporating all of the important parameters that characterize an edge/hub/cloud environment, providing an optimal allocation for a task flow graph.

\begin{figure*}[t]
    \centering
    \includegraphics[width=.85\textwidth]{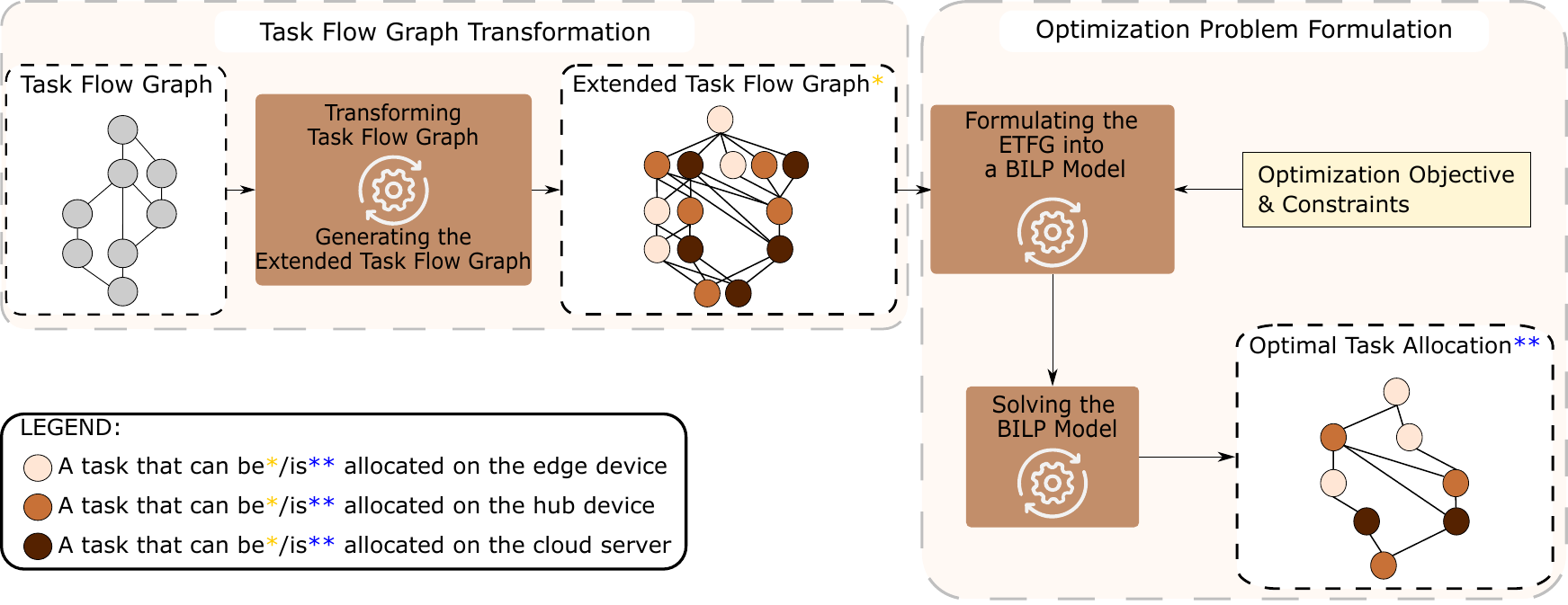}
    \caption{Overview of proposed optimization framework. The task flow graph transformation, including the encapsulated energy model, is described in \cref{extended,subsec:energyModel}. The formulation of the optimization problem is presented in \cref{subsec:optimization}.}
    \label{flow}
\end{figure*}

\subsection{Comparison with our preliminary research}
The foundational concepts of this work were first presented in a preliminary form in \cite{Kouloumpris2019}.
Below, we outline the main differences and contributions of the current study with respect to our preliminary research:
\begin{enumerate}
    \item We streamlined and enhanced the mathematical representation of all aspects of the proposed approach, from the description of the task flow graph transformation to the modeling of the optimization problem.
    
    \item We extended our optimization framework to consider a new objective for the minimization of overall energy consumption (in addition to the latency objective), based on an improved energy model.
     
    \item We developed suitable synthetic benchmarks to further validate and evaluate the efficiency and scalability of our framework, by extending our transformation method to randomly generated task flow graphs.
    
    \item We conducted extensive experimentation with alternative configurations of different devices, for both the real-world use-case scenario and the synthetic benchmarks.  
\end{enumerate}

\section{Proposed optimization framework}\label{framework}
\subsection{Framework overview}\label{frameworkOverview}
An outline of the proposed framework is illustrated in \cref{flow}. 
Given a task flow graph (TFG) of an application, we transform it into an extended task flow graph (ETFG). The ETFG represents the possible allocations of each task on the computational devices, encapsulating both the application and system models, as well as the adopted energy model (\cref{extended,subsec:energyModel}). Subsequently, the resulting ETFG is used to formulate the BILP model, by selecting the desired optimization objective, which can be either (a) the minimization of overall latency, or (b) the minimization of overall energy consumption, subject to specific constraints (\cref{subsec:optimization}). Finally, the formulated BILP problem is solved, yielding the optimal task allocation.
In the proposed  framework, we utilize BILP, as it is a suitable approach for solving problems of this nature and size at design-time (i.e., offline), guaranteeing a globally optimal solution, in contrast to heuristic methods that typically provide suboptimal results.

\subsection{Extended task flow graph (ETFG)}\label{extended}
Given an application that comprises a set of tasks $\mathcal{T} = \{1, 2, ..., |\mathcal{T}|\}$, its TFG is typically represented by a directed acyclic graph $G=( \mathcal{N}, \mathcal{A})$, where $\mathcal{N}=\{ N_{1}, N_{2}, ..., N_{|\mathcal{T}|} \}$ is the set of nodes and $\mathcal{A} = \{ A_{i \rightarrow j} \, | \, N_{i}, N_{j} \in \mathcal{N}, \, i \neq j, \, \exists \text{ a data dependency } N_{i} \rightarrow N_{j} \}$ is the set of arcs in the graph. A node $N_{i} \in \mathcal{N}$ corresponds to a task $i \in \mathcal{T}$ of the application. An arc $A_{i \rightarrow j} \in \mathcal{A}$ represents the data flow between nodes $N_{i}$ and $N_{j}$, indicating the precedence relationship and communication between parent task $i$ and child task $j$. 

The initial TFG $G$ is transformed into the ETFG $G^{\prime}=(\mathcal{N}^{\prime}, \mathcal{A}^{\prime})$ as follows. 
First, each node $N_{i} \in \mathcal{N}$ in $G$  is transformed into a composite node (i.e., a set of nodes) $N_{i}^{\prime} \in \mathcal{N}^{\prime}$ in $G^{\prime}$, such that:
\begin{equation}
\label{eq:nodeTransformation}
 f_{\mathrm{node}}(N_{i}) = N_{i}^{\prime} = \{ N_{ik} \, | \, k \in \mathcal{F}_{i} \subseteq \mathcal{U} \},
\end{equation}
where $f_{\mathrm{node}}$ is the function that transforms node $N_{i}$ into the set $N_{i}^{\prime}$ as shown above, whereas $k$ is a device in  $\mathcal{F}_{i} \subseteq \mathcal{U}$. 
$\mathcal{U} = \{ \mathrm{e}, \mathrm{h}, \mathrm{c} \}$ is the set of computational devices in the target edge/hub/cloud environment, where $\mathrm{e}$ is an edge device, $\mathrm{h}$ is a hub device, and $\mathrm{c}$ is a cloud server. $\mathcal{F}_{i}$ is the subset of devices where task $i$ can be allocated.
If task $i$ requires fixed allocation on specific devices in $\mathcal{U}$, then $\mathcal{F}_{i} \subset \mathcal{U}$, otherwise $\mathcal{F}_{i} = \mathcal{U}$.
Consequently, an individual node $N_{ik} \in G^{\prime}$ indicates the possible allocation of task $i$ on device $k$.

Subsequently, each arc $A_{i \rightarrow j} \in \mathcal{A}$ in $G$ is transformed into a composite arc (i.e., a set of arcs) $A^{\prime}_{i \rightarrow j} \in \mathcal{A}^{\prime}$ in $G^{\prime}$, such that:
\begin{equation}
 \label{eq:arcTransformation}
 \begin{split}
     f_{\mathrm{arc}}(A_{i \rightarrow j}) = A^{\prime}_{i \rightarrow j}  =  \{ A_{ik \rightarrow jl}= N_{ik} \rightarrow N_{jl} \, | & \, k \in \mathcal{F}_{i} \subseteq \mathcal{U} , \\
     & \, l \in \mathcal{F}_{j} \subseteq \mathcal{U}  \},
\end{split}
\end{equation}
where $f_{\mathrm{arc}}$ is the function that transforms arc $A_{i \rightarrow j}$ into the set $A^{\prime}_{i \rightarrow j}$ as shown above. $k$ and $l$ denote devices where tasks $i$ and $j$ can be allocated, respectively.
Hence, an individual arc $A_{ik \rightarrow jl} \in G^{\prime}$ indicates the data flow between nodes $N_{ik}$ and $N_{jl}$. If both tasks $i$ and $j$ are allocated on the same device (i.e., $k=l$), the cost for their communication (in terms of latency and energy) is considered negligible, compared to the case where $k \neq l$. If $k \neq l$ and there is no direct communication channel between devices $k$ and $l$ (i.e., between the edge device $\mathrm{e}$ and the cloud server $\mathrm{c}$), an additional communication cost is considered, as the communication must be conducted through an intermediate device $m$ (i.e., the hub device $\mathrm{h}$).

A node $N_{ik} \in G^{\prime}$ is also referred to as a \emph{candidate node}. Only one candidate node $N_{ik}$ in a composite node $N_{i}^{\prime}$ will eventually be selected in the solution of the problem to allocate task $i$ on a specific device $k$. Based on the selected candidate nodes, the corresponding arcs in $G^{\prime}$ will be selected as well. The selection of the candidate nodes and their corresponding arcs will be performed by the BILP solver, based on the desired objective and the constraints defined in \cref{subsec:optimization}.

\subsubsection{Parameters of ETFG candidate nodes}\label{subsec:paramsNodes}
A candidate node $N_{ik} \in G^{\prime}$ has the following parameters, which reflect the resource requirements and characteristics of task $i$ when allocated on device $k$: 
\begin{enumerate}
    \item \emph{Computational latency} $L_{ik}$: execution time of task $i$ on device $k$.
   
    \item \emph{Computational power consumption} $P_{ik}$: power required to execute task $i$ on device $k$. 
    
    \item \emph{Computational energy consumption} $E_{ik}$: energy required to execute task $i$ on device $k$.
     
    \item \emph{Memory} $M_{i}$: main memory required by task $i$.
    
    \item \emph{Storage} $S_{i}$: storage required by task $i$.
    
    \item \emph{Output data} $D_{i}$: size of output data generated by task $i$.
    
    \item \emph{Number of child tasks} $NC_{i}$: number of child tasks of  task $i$ (i.e., number of immediate successors of node $N_{i} \in G$).  
\end{enumerate}

The parameters $L_{ik}$, $P_{ik}$, and $E_{ik}$ are device-dependent, whereas $M_{i}$, $S_{i}$, $D_{i}$, and $NC_{i}$ are  device-independent. 
$E_{ik}$ is calculated using $P_{ik}$ and $L_{ik}$, based on the energy model presented in \cref{subsec:energyModel}.
The parameters $L_{ik}$, $P_{ik}$, $M_{i}$, $S_{i}$, and $D_{i}$ can be determined via performance and power profiling tools and instruments, as described in \cref{evaluation}. On the other hand, $NC_{i}$ is derived from the structure of TFG $G$.

\subsubsection{Parameters of ETFG arcs}\label{subsec:paramsArcs}
An arc $A_{ik \rightarrow jl} \in G^{\prime}$ has the following parameters:
\begin{enumerate}
    \item \emph{Communication latency} $CL_{ik \rightarrow jl}$: time required to transfer the output data $D_{i}$ of parent task $i$, allocated on device $k$, to child task $j$,  allocated on device $l$. 
   
    \item \emph{Communication energy consumption} $CE_{ik \rightarrow jl}$: energy required to transfer $D_{i}$ from task $i$ to task $j$, allocated on devices $k$ and $l$, respectively.
    
    \item \emph{Indirect communication indicator} $\delta_{ik \rightarrow jl}^{m}$: binary parameter denoting whether arc $A_{ik \rightarrow jl}$ involves indirect communication between devices $k$ and $l$ via  intermediate device $m$. It is defined as:
    \begin{equation}\label{eq:indicator}
        \delta_{ik \rightarrow jl}^{m} = 
        \begin{cases}
            1, & \text{if $(k,l) \in \left\{ (\mathrm{e}, \mathrm{c}), (\mathrm{c}, \mathrm{e}) \right\}, \, m=\mathrm{h}$ },\\
            0, & \text{otherwise}.
        \end{cases}
    \end{equation}  
\end{enumerate}

The communication latency $CL_{ik \rightarrow jl}$ depends on the output data size $D_{i}$ and the bandwidth of the utilized communication channel. It is given by:
\begin{equation}\label{eq:commLatency}
     CL_{ik \rightarrow jl} = 
     \begin{cases}
        D_{i} / W_{kl}, & \text{if $\delta_{ik \rightarrow jl}^{m}=0, \, k \neq l$},\\
        D_{i}  \left( 1 / W_{km} + 1 / W_{ml} \right), & \text{if $\delta_{ik \rightarrow jl}^{m}=1$},\\
        0, & \text{if $k = l$},
     \end{cases}
\end{equation}
where $W_{kl}$, $W_{km}$, and $W_{ml}$ denote the bandwidth of the communication channels between the corresponding devices.
Similarly, the communication energy consumption $CE_{ik \rightarrow jl}$ depends on $D_{i}$ and the energy required to transmit and receive a unit of data over the utilized communication channel, as explained in \cref{subsec:energyModel}.

\subsubsection{TFG to ETFG transformation example}\label{example}
\begin{figure}[!t]
    \centering
    \includegraphics[width=\columnwidth]{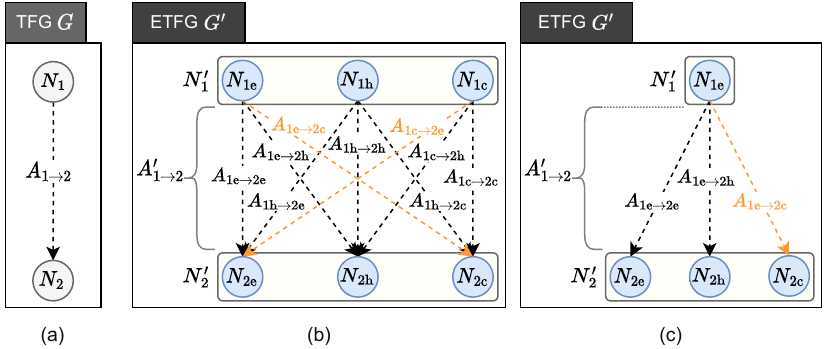}
    \caption{Example of transforming a TFG $G$ into ETFG $G^{\prime}$, considering two different cases of task allocation requirements.}
    \label{etfg_diagram}
\end{figure}

The transformation of a TFG $G$ into ETFG $G^{\prime}$ is illustrated in \cref{etfg_diagram}, under two different scenarios of task allocation requirements.
\cref{etfg_diagram}a depicts $G$, which consists of nodes $N_1$ and $N_2$ (corresponding to tasks 1 and 2, respectively). 
\cref{etfg_diagram}b shows the transformation of $G$ into $G^{\prime}$ in the case where both tasks can be allocated on any device, i.e., $\mathcal{F}_1 \allowbreak = \allowbreak \mathcal{F}_2 \allowbreak = \allowbreak \mathcal{U} \allowbreak = \allowbreak \{\mathrm{e}, \allowbreak \mathrm{h}, \allowbreak \mathrm{c} \}$. 
First, the nodes $N_1, N_2 \in G$ are transformed into the composite nodes (i.e., sets of candidate nodes) $N_1^{\prime}, N_2^{\prime} \in G^{\prime}$, respectively. As both tasks can be allocated on any device, a candidate node is generated for each task-device combination, i.e., $N_1^{\prime} \allowbreak = \{ N_{1\mathrm{e}}, \allowbreak N_{1\mathrm{h}}, \allowbreak N_{1\mathrm{c}} \}$ and $N_2^{\prime} = \allowbreak \{ N_{2\mathrm{e}}, \allowbreak N_{2\mathrm{h}}, \allowbreak N_{2\mathrm{c}} \}$. Subsequently, the arc $A_{1 \rightarrow 2} \in G$ is transformed into the composite arc (i.e., set of arcs) $A_{1 \rightarrow 2}^{\prime} \in G^{\prime}$. In $A_{1 \rightarrow 2}^{\prime}$, an individual arc  is generated between each pair of candidate nodes of the two tasks. Thus, $A_{1 \rightarrow 2}^{\prime} \allowbreak = \{ A_{1\mathrm{e} \rightarrow 2\mathrm{e}}, \allowbreak A_{1\mathrm{e} \rightarrow 2\mathrm{h}}, \allowbreak A_{1\mathrm{e} \rightarrow 2\mathrm{c}}, \allowbreak A_{1\mathrm{h} \rightarrow 2\mathrm{e}}, \allowbreak A_{1\mathrm{h} \rightarrow 2\mathrm{h}}, \allowbreak A_{1\mathrm{h} \rightarrow 2\mathrm{c}}, \allowbreak A_{1\mathrm{c} \rightarrow 2\mathrm{e}}, \allowbreak A_{1\mathrm{c} \rightarrow 2\mathrm{h}}, \allowbreak A_{1\mathrm{c} \rightarrow 2\mathrm{c}} \}$. The arcs $A_{1\mathrm{e} \rightarrow 2\mathrm{c}}$ and $A_{1\mathrm{c} \rightarrow 2\mathrm{e}}$ involve indirect communication between devices $\mathrm{e}$ and $\mathrm{c}$, through device $\mathrm{h}$. To denote this, they are depicted in orange.

\cref{etfg_diagram}c illustrates the transformation of $G$ into $G^{\prime}$ in the case where task 1 requires fixed allocation on device $\mathrm{e}$ (i.e., $\mathcal{F}_1 = \{ \mathrm{e} \}$), whereas task 2 can be allocated on any device (i.e., $\mathcal{F}_2 \allowbreak = \allowbreak \{\mathrm{e}, \allowbreak \mathrm{h}, \allowbreak \mathrm{c} \}$). The nodes $N_1, N_2 \in G$ are transformed into the composite nodes $N_1^{\prime}, N_2^{\prime} \in G^{\prime}$, respectively. In this scenario, as task 1 can be allocated only on device $\mathrm{e}$, $N_1^{\prime} \allowbreak = \{ N_{1\mathrm{e}}\}$. On the other hand, $N_2^{\prime} = \allowbreak \{ N_{2\mathrm{e}}, \allowbreak N_{2\mathrm{h}}, \allowbreak N_{2\mathrm{c}} \}$, as in the previous scenario (\cref{etfg_diagram}b).
The arc $A_{1 \rightarrow 2} \in G$ is transformed into the composite arc $A_{1 \rightarrow 2}^{\prime} \in G^{\prime}$, such that $A_{1 \rightarrow 2}^{\prime} \allowbreak = \{ A_{1\mathrm{e} \rightarrow 2\mathrm{e}}, \allowbreak A_{1\mathrm{e} \rightarrow 2\mathrm{h}}, \allowbreak A_{1\mathrm{e} \rightarrow 2\mathrm{c}} \}$. Similar to \cref{etfg_diagram}b, $A_{1\mathrm{e} \rightarrow 2\mathrm{c}}$ is shown in orange, as it involves indirect communication between devices $\mathrm{e}$ and $\mathrm{c}$.

The proposed transformation of TFG $G$ into ETFG $G^{\prime}$ increases the size of the graph. In the worst case, where no task requires fixed allocation, for $\nu$ different devices, the number of nodes in $G^{\prime}$ increases by $\nu$ times, whereas the number of arcs increases by $\nu^2$ times, with respect to those in $G$. As the targeted applications require only three devices, the benefits of the proposed framework outweigh the added complexity due to the increase in the size of $G^{\prime}$. We showcase this experimentally in \cref{evaluation}.

\subsection{Energy model}\label{subsec:energyModel}
The energy model is encapsulated in the ETFG $G^{\prime}$ through the parameters $E_{ik}$ and $CE_{ik \rightarrow jl}$, considering both the computational and communication energy requirements of the application \cite{Wang2017}.
Specifically, the computational energy consumption $E_{ik}$ required to execute task $i$ on device $k$ is defined as:
\begin{equation}\label{eq:compEnergy}
    E_{ik} = P_{ik} \,  L_{ik}.
\end{equation}
On the other hand, the communication energy consumption $CE_{ik \rightarrow jl}$ required to transfer the output data $D_{i}$ of parent task $i$, allocated on device $k$, to child task $j$, allocated on device $l$, is given by:
\begin{equation}\label{eq:commEnergy}
    CE_{ik \rightarrow jl} = 
    \begin{cases}
        D_{i}  \left(\tau_{kl} + \rho_{kl} \right), & \text{if $\delta_{ik \rightarrow jl}^{m}=0, \, k \neq l$},\\
        D_{i}  \left(\tau_{km} + \rho_{km} + \tau_{ml} + \rho_{ml}  \right), & \text{if $\delta_{ik \rightarrow jl}^{m}=1$},\\
        0, & \text{if $k = l$}.
    \end{cases}
\end{equation}
The parameters $\tau_{kl}$, $\tau_{km}$, and $\tau_{ml}$ indicate the energy required to transmit a unit of data, whereas $\rho_{kl}$, $\rho_{km}$, and $\rho_{ml}$ denote the energy required to receive a unit of data, over the respective communication channels.

\subsection{Optimization problem formulation}\label{subsec:optimization}
The ETFG $G^{\prime}$ is utilized to formulate the considered task allocation problem as a BILP model.
Depending on the use-case, our framework minimizes the desired objective, either the overall latency or overall energy consumption, by selecting the appropriate candidate nodes and arcs in $G^{\prime}$, based on the following formulation.

\subsubsection{Decision variables}
We employ the following binary decision variables:
\begin{enumerate}
    \item $x_{ik}$: corresponds to a candidate node $N_{ik} \in G^{\prime}$, such that $x_{ik} = 1$ if $N_{ik}$ is selected (i.e., task $i$ is allocated on device $k$), and $x_{ik} = 0$ otherwise.
   
    \item $x_{ik \rightarrow jl}$: corresponds to an arc $A_{ik \rightarrow jl} \in G^{\prime}$, such that $x_{ik \rightarrow jl} = 1$ if $A_{ik \rightarrow jl}$ is selected, and $x_{ik \rightarrow jl} = 0$ otherwise.
\end{enumerate}

\subsubsection{Objective function 1}
The first objective concerns the minimization of overall latency:
\begin{equation}
\label{equ:objective}
\min \left( \sum_{i \in \mathcal{T}} \sum_{k \in \mathcal{F}_{i}} L_{ik} \, x_{ik} +  \sum_{i \in \mathcal{T}} \sum_{k \in \mathcal{F}_{i}} \sum_{j \in \mathcal{T}} \sum_{l \in \mathcal{F}_{j}} CL_{ik \rightarrow jl} \, x_{ik \rightarrow jl} \right).
\end{equation}
The first term in \eqref{equ:objective} represents the total computational latency, whereas the second term denotes the total communication latency.

\subsubsection{Objective function 2}
The second objective pertains to the minimization of overall energy consumption:
\begin{equation}
\label{equ:objective2}
\min \left( \sum_{i \in \mathcal{T}} \sum_{k \in \mathcal{F}_{i}} E_{ik} \, x_{ik} +  \sum_{i \in \mathcal{T}} \sum_{k \in \mathcal{F}_{i}} \sum_{j \in \mathcal{T}} \sum_{l \in \mathcal{F}_{j}} CE_{ik \rightarrow jl} \, x_{ik \rightarrow jl} \right).
\end{equation}
Similar to the latency objective, the first term in \eqref{equ:objective2} indicates the total computational energy consumption, whereas the second term represents the total communication energy consumption.

\subsubsection{Constraints}\label{subsubsec:constraints}
The desired objective function, \eqref{equ:objective} or \eqref{equ:objective2}, is solved subject to the following constraints:
\begin{equation}
    \label{eq:one}
     x_{ik} \in \{ 0, 1 \}, \, \forall \, i \in \mathcal{T}, \, \forall \, k \in \mathcal{F}_{i},  
\end{equation}
\begin{equation}
    \label{eq:two}
     x_{ik \rightarrow jl} \in \{ 0, 1 \}, \, \forall \, i, j \in \mathcal{T}, \, \forall \, k \in \mathcal{F}_{i}, \, \forall \, l \in \mathcal{F}_{j}, \, i \neq j,   
\end{equation}
\begin{equation}
    \label{eq:three}
    \sum_{k \in \mathcal{F}_{i}} x_{ik} = 1, \, \forall \, i \in \mathcal{T},  
\end{equation}
\begin{equation}
    \label{eq:four}
    \sum_{k \in \mathcal{F}_{i}} \sum_{j \in \mathcal{T}} \sum_{l \in \mathcal{F}_{j}} x_{ik \rightarrow jl} = NC_{i}, \, \forall \, i \in \mathcal{T}, \, i \neq j,
\end{equation}
\begin{equation}
    \label{eq:five1}
    x_{ik \rightarrow jl} \leq x_{ik}, \, \forall \, i, j \in \mathcal{T}, \, \forall \, k \in \mathcal{F}_{i}, \, \forall \, l \in \mathcal{F}_{j}, \, i \neq j,
\end{equation}
\begin{equation}
    \label{eq:five2}
    x_{ik \rightarrow jl} \leq x_{jl}, \, \forall \, i, j \in \mathcal{T}, \, \forall \, k \in \mathcal{F}_{i}, \, \forall \, l \in \mathcal{F}_{j}, \, i \neq j,
\end{equation}
\begin{equation}
    \label{eq:five3}
    x_{ik \rightarrow jl} \geq x_{ik} + x_{jl} - 1, \, \forall \, i, j \in \mathcal{T}, \, \forall \, k \in \mathcal{F}_{i}, \, \forall \, l \in \mathcal{F}_{j}, \, i \neq j,
\end{equation}
\begin{equation}
    \label{eq:six}
    \sum_{i \in \mathcal{T}} M_{i} \, x_{ik} \leq M_{k}^{\mathrm{bgt}}, \, \forall \, k \in \mathcal{U},
\end{equation}
\begin{equation}
    \label{eq:seven}
    \sum_{i \in \mathcal{T}} S_{i} \, x_{ik} \leq S_{k}^{\mathrm{bgt}}, \, \forall \, k \in \mathcal{U},
\end{equation}
\begin{equation}
    \label{eq:eight}
    \begin{split}
    & \sum_{i \in \mathcal{T}} E_{ik} \, x_{ik} \\
    & + \sum_{i \in \mathcal{T}} \sum_{j \in \mathcal{T}} \sum_{l \in \mathcal{F}_{j}} \sum_{m \in \mathcal{U}} \bigg( 
    D_{i} \, x_{ik \rightarrow jl} \left( \tau_{kl} \left( 1 - \delta_{ik \rightarrow jl}^{m} \right) + \tau_{km} \, \delta_{ik \rightarrow jl}^{m} \right)\\
    & + D_{j} \, x_{jl \rightarrow ik} \left( \rho_{lk} \left( 1 - \delta_{jl \rightarrow ik}^{m} \right) + \rho_{mk} \, \delta_{jl \rightarrow ik}^{m} \right) \bigg)\\
    & + \sum_{i \in \mathcal{T}} \sum_{l \in \mathcal{F}_{i}} \sum_{j \in \mathcal{T}} \sum_{m \in \mathcal{F}_{j}} D_{i} \, x_{il \rightarrow jm} \left( \rho_{lk} + \tau_{km} \right)  \delta_{il \rightarrow jm}^{k}\\
    & \leq  E_{k}^{\mathrm{bgt}}, \, \forall \, k \in \mathcal{U}, \, i \neq j, \, k \neq l \neq m.
    \end{split}
\end{equation}

In the case of objective function \eqref{equ:objective2} (energy objective), the following additional constraint is considered:
\begin{equation}
    \label{eq:nine}
    \sum_{i \in \mathcal{T}} \sum_{k \in \mathcal{F}_{i}} L_{ik} \, x_{ik} +  \sum_{i \in \mathcal{T}} \sum_{k \in \mathcal{F}_{i}} \sum_{j \in \mathcal{T}} \sum_{l \in \mathcal{F}_{j}} CL_{ik \rightarrow jl} \, x_{ik \rightarrow jl} \leq  L_{\mathrm{thr}}, \, i \neq j.
\end{equation}

Constraints \eqref{eq:one} and \eqref{eq:two} ensure the binary nature of the decision variables $x_{ik}$ and $x_{ik \rightarrow jl}$. 
Constraint \eqref{eq:three} guarantees that only one candidate node $N_{ik}$ will be selected for each task $i \in \mathcal{T}$ (i.e., it ensures that each task $i \in \mathcal{T}$ will be allocated on only one device $k \in \mathcal{F}_{i}$). 
On the other hand, \eqref{eq:four} guarantees that for each candidate node $N_{ik}$ the number of outgoing arcs that will be selected will be equal to the number of child tasks $NC_{i}$ of task $i$. In the same context, \eqref{eq:five1}--\eqref{eq:five3} ensure that if candidate nodes $N_{ik}$ and $N_{jl}$ are selected, the corresponding arc $A_{ik \rightarrow jl}$ will be selected as well.
Thus, \eqref{eq:four}--\eqref{eq:five3} guarantee that the connectivity and precedence relationships between the tasks will be preserved. 
Constraints \eqref{eq:six}--\eqref{eq:eight} ensure that the memory $M_{k}^{\mathrm{bgt}}$, storage $S_{k}^{\mathrm{bgt}}$, and energy $E_{k}^{\mathrm{bgt}}$ budgets, respectively, for each device $k \in \mathcal{U}$ will not be exceeded for the execution of the application. These budgets are defined based on the resource limitations of each device.
Finally, if the desired objective is the minimization of energy, \eqref{eq:nine} guarantees that a predefined latency threshold $L_{\mathrm{thr}}$ that bounds the acceptable overall latency is not exceeded.

It is noted that \eqref{eq:eight} is considered not only when minimizing latency, but also when minimizing energy, as even in the latter case, the energy budget of one or more devices may be exceeded.
Moreover, it is noted that both the computational and communication energy consumption of a device $k$ are taken into account in \eqref{eq:eight}. With respect to the communication energy, we consider all the data transmitted from and received at device $k$ (either directly or indirectly to/from another device), including the case where $k$ is used for the communication between other devices.
The notations used in the proposed framework are included in Table A.1 in Appendix A.

\section{Experimental analysis and results}\label{evaluation}

\begin{table*}[t]
\centering
\caption{Computational devices.}
\footnotesize
\begin{tabular}{@{\extracolsep{4pt}}cccrrr@{}}
\toprule
\multirow{2}{*}{Device $k$} & \multirow{2}{*}{$\mathrm{e}$/$\mathrm{h}$/$\mathrm{c}$} & \multirow{2}{*}{Processor} & \multicolumn{3}{c}{Budgets} \\
\cline{4-6}
  & & & \multicolumn{1}{c}{Memory $M_{k}^{\mathrm{bgt}}$} & \multicolumn{1}{c}{Storage $S_{k}^{\mathrm{bgt}}$} & Energy $E_{k}^{\mathrm{bgt}}$\\
\hline
Jetson TX2 & $\mathrm{e}$ & Cortex-A57\,/\,NVIDIA Denver2 @ 2\,GHz &  8\,GiB \hspace{10pt}   & 32\,GiB \hspace{6pt}   & 129.96\,Wh*\\
Odroid XU4 & $\mathrm{e}$ & Exynos 5422 Cortex-A15\,/\,Cortex-A7 @ 2\,GHz       & 2\,GiB \hspace{10pt}   & 16\,GiB \hspace{6pt}    & 129.96\,Wh*\\
Raspberry Pi 3 Model B & $\mathrm{e}$ & Broadcom Cortex-A53 @ 1.4\,GHz         & 1\,GiB \hspace{10pt}   & 16\,GiB \hspace{6pt}    & 129.96\,Wh*\\
Mi Notebook Pro & $\mathrm{h}$ & Intel i5 8250U @ 1.6\,GHz                     & 8\,GiB \hspace{10pt}   & 512\,GiB \hspace{6pt}   & 60.00\,Wh\dag\\
HPE ProLiant  DL580 Gen10 & $\mathrm{c}$ & Intel Xeon Gold 6240 @ 2.6\,GHz     & 400\,GiB \hspace{10pt} & 10\,TiB \hspace{6pt}    & \multicolumn{1}{c}{-}\\

\bottomrule

\multicolumn{6}{l}{\scriptsize{*Battery capacity of UAV where device is attached as payload. \hspace{2mm} \dag Battery capacity of device.}}\\

\end{tabular}
\label{constraints}
\end{table*}

\begin{table}[t]
\centering
\caption{Configurations of computational devices.}
\footnotesize
\begin{tabular}{@{\extracolsep{4pt}}ccccc@{}}
\toprule
\multirow{2}{*}{Device} & \multirow{2}{*}{$\mathrm{e}$/$\mathrm{h}$/$\mathrm{c}$} & \multicolumn{3}{c}{Configuration} \\
\cline{3-5}
 & & C1 & C2 & C3\\ 
\hline
Jetson TX2 & $\mathrm{e}$ & \checkmark & &  \\

Odroid XU4 & $\mathrm{e}$ &  & \checkmark & \\

Raspberry Pi 3 Model B & $\mathrm{e}$ & & & \checkmark \\

Mi Notebook Pro & $\mathrm{h}$ & \checkmark & \checkmark & \checkmark\\

HPE ProLiant DL580 Gen10 & $\mathrm{c}$ & \checkmark & \checkmark & \checkmark\\

\bottomrule

\end{tabular}
\label{configurations}
\end{table}

\subsection{Evaluation strategy}
\label{subsec:strategy}
We investigated our approach by conducting an extensive number of experiments:
\begin{enumerate}
    \item First, we validated our method utilizing a relevant real-world use-case scenario for the UAV-based inspection of power transmission towers and lines.
    
    \item Subsequently, we further examined the efficiency and scalability of our framework by developing and using appropriate synthetic benchmarks.
\end{enumerate}

We explored three alternative configurations (combinations) of the edge device $\mathrm{e}$, the hub device $\mathrm{h}$, and the cloud server $\mathrm{c}$. In each configuration, a different device $\mathrm{e}$ was considered.
For the first set of experiments, we examined two distinct objectives, the minimization of either latency or energy, performing two different runs for each configuration. In each run, we took into account varied communication channel characteristics.
As our primary goal in the second set of experiments was to evaluate the scalability of the proposed approach, without loss of generality, we examined one of the optimization objectives, the minimization of latency, conducting a run for each configuration and synthetic benchmark.
In contrast to the qualitative comparison with existing methods presented in \Cref{background}, a meaningful quantitative comparative evaluation does not apply, as existing approaches do not include all the parameters of our framework, do not provide the optimal solution, or do not use the underlying architecture considered in this work.

\subsection{Experimental setup}\label{setup}
Devices $\mathrm{e}$, $\mathrm{h}$, and $\mathrm{c}$ were based on real-world counterparts typically used in emerging applications, such as the ones motivating this work. Specifically, as device $\mathrm{e}$ we selected either Jetson TX2, Odroid XU4, or Raspberry Pi 3 Model B to represent a high, middle, or low-end edge device, respectively. Since these devices can be attached to a UAV as payload to provide computational capability, for battery capacity purposes, we considered a DJI Matrice 100 UAV.
We used Mi Notebook Pro as device $\mathrm{h}$, as it can facilitate the communication between devices $\mathrm{e}$ and $\mathrm{c}$.
Finally, the HPE ProLiant DL580 Gen10 server was selected as device $\mathrm{c}$, since its computational capacity is representative of that provided by a high-end server (physical or virtual) in a cloud environment. \cref{constraints} shows the devices we considered in our experiments, along with their corresponding memory, storage, and energy budgets. \cref{configurations} demonstrates the three alternative device configurations we investigated (C1, C2, and C3). The check marks indicate the devices included in each configuration.

\cref{table:energy_table} shows the $W_{kl}$, $\tau_{kl}$, and $\rho_{kl}$ parameters we used for each communication channel per run (Run 1 and Run 2). The particular values were selected to reflect the variations in bandwidth and energy consumption that may be observed due to the distance between the devices and the terrain in which the application is deployed (e.g., rural or urban terrain) \cite{4GMobile, 4Genergyconsumption}. 
In the real-world use-case scenario, for each configuration C1, C2, and C3 we conducted two runs, one for the bandwidth and energy parameters of Run 1, and one for those of Run 2. 
Moreover, for the energy objective,  the latency threshold was set to $L_{\mathrm{thr}}=8000$\,ms, which is a representative, but still challenging, time constraint for the particular application. 
On the other hand, in the case of synthetic benchmarks, for each configuration C1, C2, and C3 we performed a run using the bandwidth and energy parameters of Run 1.
The TFG transformation and the optimization problem formulation were implemented in C++. The formulated problem in each case was solved using Gurobi Optimizer 9.5 \cite{gurobi}, run on a server equipped with an Intel Xeon Gold 6240 processor @ 2.6\,GHz and 367\,GiB of RAM.

\begin{table*}[t]
\centering
\caption{Bandwidth and energy parameters for each communication channel per run.}
\footnotesize
\begin{tabular}{@{\extracolsep{4pt}}ccccrcc@{}} 
\toprule
Communication Channel & \multicolumn{3}{c}{Run 1} & \multicolumn{3}{c}{Run 2}\\
\cline{2-4} \cline{5-7}
Between Devices $k \rightarrow l$ & $W_{kl}$ (Mbit/s) & $\tau_{kl}$ (\SI{}{\micro \joule}/bit) & $\rho_{kl}$ (\SI{}{\micro \joule}/bit) & $W_{kl}$ (Mbit/s) & $\tau_{kl}$ (\SI{}{\micro \joule}/bit) & $\rho_{kl}$ (\SI{}{\micro \joule}/bit)\\ 
 \hline
 $\mathrm{e} \rightarrow \mathrm{h}$ & 15 & 1.0 & 0.70 & 10.0 \hspace{10pt} & 1.0 & 0.7 \\
 $\mathrm{h} \rightarrow \mathrm{e}$ & 20 & 1.0 & 0.70 & 10.0 \hspace{10pt} & 1.0 & 0.7 \\
 $\mathrm{h} \rightarrow \mathrm{c}$ & 25 & 2.5 & 1.25 & 0.5 \hspace{10pt} & 6.5 & 4.5 \\
 $\mathrm{c} \rightarrow \mathrm{h}$ & 35 & 2.5 & 1.25 & 1.5 \hspace{10pt} & 6.5 & 4.5 \\
\bottomrule
\end{tabular}
\label{table:energy_table}
\end{table*}

\subsection{Real-world use-case scenario}\label{case_study}
\subsubsection{Overview of real-world application}\label{real_app}
The real-world application \cite{icarusTheocharides}, which concerns the UAV-based aerial visual inspection of power transmission towers and lines, consists of 15 tasks. Its flowchart, encapsulating its TFG, as well as its corresponding ETFG, are depicted in Fig. A.1 in Appendix A. 
A brief description of its tasks is given below:
\begin{enumerate}
    \item Image acquisition (task 1): captures an image through a camera connected to the edge device on the UAV.
    
    \item Image preprocessing (tasks 2--5): preliminary processing of the captured image.
    
    \item Power transmission lines detection (tasks 6--9): Hough transform-based line detector that identifies the transmission lines for navigation purposes, while detecting transmission line anomalies, such as vegetation.
    
    \item Power transmission towers detection (tasks 10--14): convolutional neural network (CNN) based detector that registers the location of transmission towers and inspects them for problems, such as faulty insulators and spacers.
    
    \item Display results (task 15): displays the final output on the hub device used by the operator to control the UAV.
\end{enumerate}

Tasks 1 and 15 can be allocated only on devices $\mathrm{e}$ and $\mathrm{h}$, respectively. On the other hand, tasks 2--14 can be allocated on any device, $\mathrm{e}$, $\mathrm{h}$, or $\mathrm{c}$. 
This is reflected in the ETFG in Fig. A.1b (Appendix A), where only one candidate node was generated for tasks 1 and 15, whereas three candidate nodes were generated for each of the other tasks.
The parameters $L_{ik}$, $P_{ik}$, $M_{i}$, $S_{i}$, and $D_{i}$ of each candidate node $N_{ik}$ in the ETFG were determined by profiling the execution of task $i$ on device $k$, in each configuration. For this purpose, we used  performance and power profiling tools (Sysprof, perf, and PowerTOP), as well as a digital power monitoring device \cite{sysprof, powertop}. 
$E_{ik}$ was calculated using \eqref{eq:compEnergy}, based on $P_{ik}$ and $L_{ik}$.
$NC_{i}$ was derived from the structure of the TFG.
On the other hand, $\delta_{ik \rightarrow jl}^{m}$ was derived from the structure of the ETFG. 
The parameters $CL_{ik \rightarrow jl}$ and $CE_{ik \rightarrow jl}$ were calculated using \eqref{eq:commLatency} and \eqref{eq:commEnergy}, respectively, based on $D_{i}$ and the bandwidth and energy parameters defined for Run 1 and Run 2 in \cref{table:energy_table}.

\subsubsection{Framework evaluation -- Real-world use-case scenario}\label{case_study-results}
\begin{figure*}[t]
    \centering
    \begin{subfigure}{.9\textwidth}
        \includegraphics[width=\columnwidth]{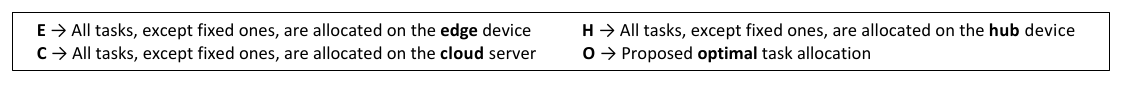}%
    \end{subfigure}\hfill%
    \begin{subfigure}[b]{.495\textwidth}
        \includegraphics[width=\columnwidth, height=1.8in]{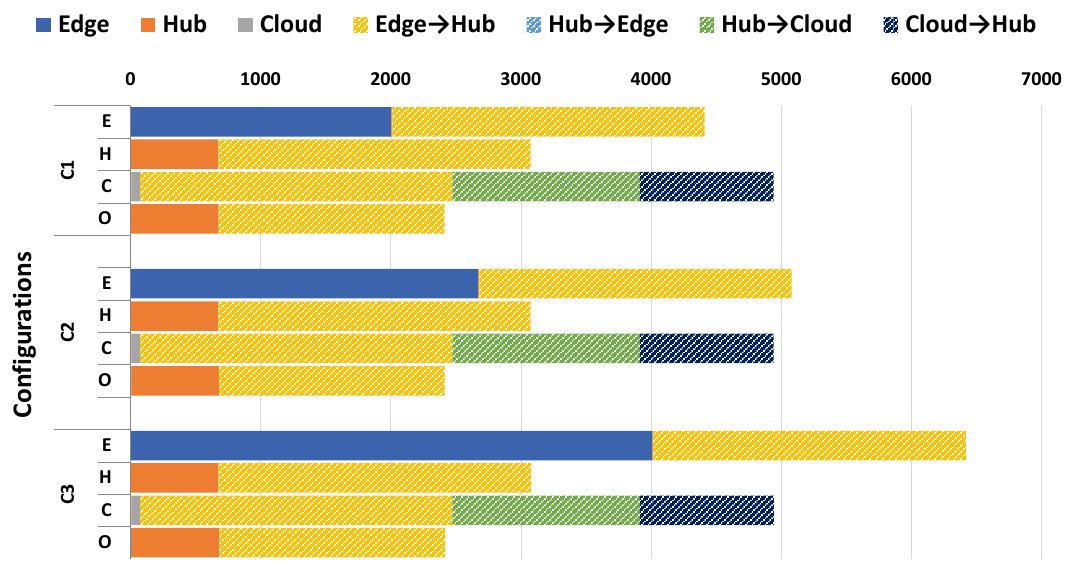}%
        \caption{Latency (in ms) per configuration (Run 1)}%
        \label{results1}%
    \end{subfigure}\hfill%
    \begin{subfigure}[b]{.495\textwidth}
        \includegraphics[width=\columnwidth, height=1.8in]{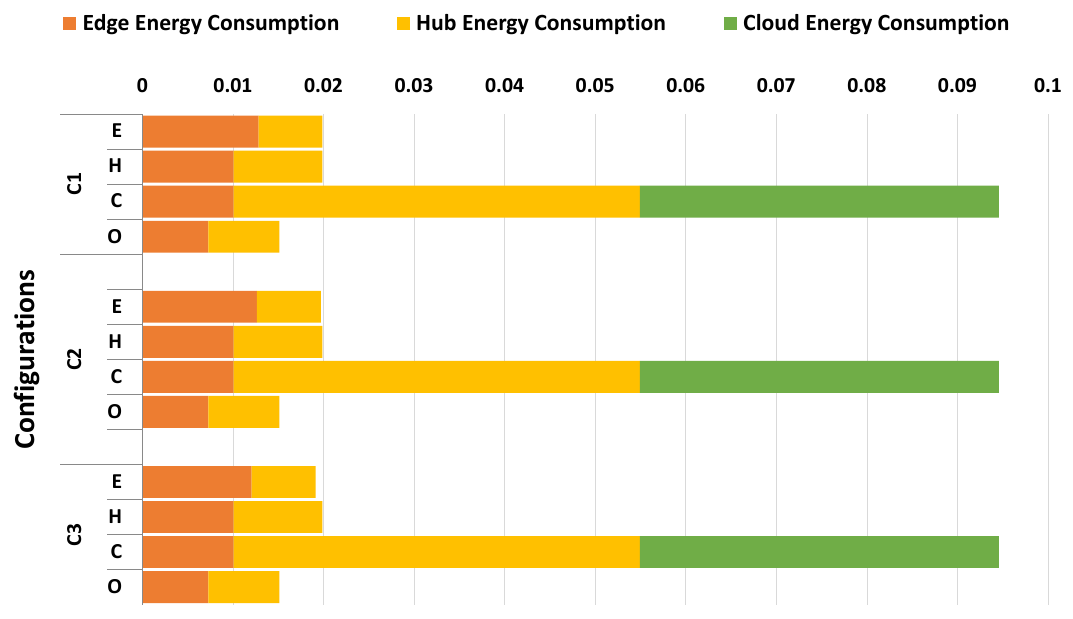}%
        \caption{Energy consumption (in Wh) per configuration (Run 1)}%
        \label{results2}%
    \end{subfigure}\hfill 
    \begin{subfigure}[b]{.495\textwidth}
        \includegraphics[width=\columnwidth, height=1.8in]{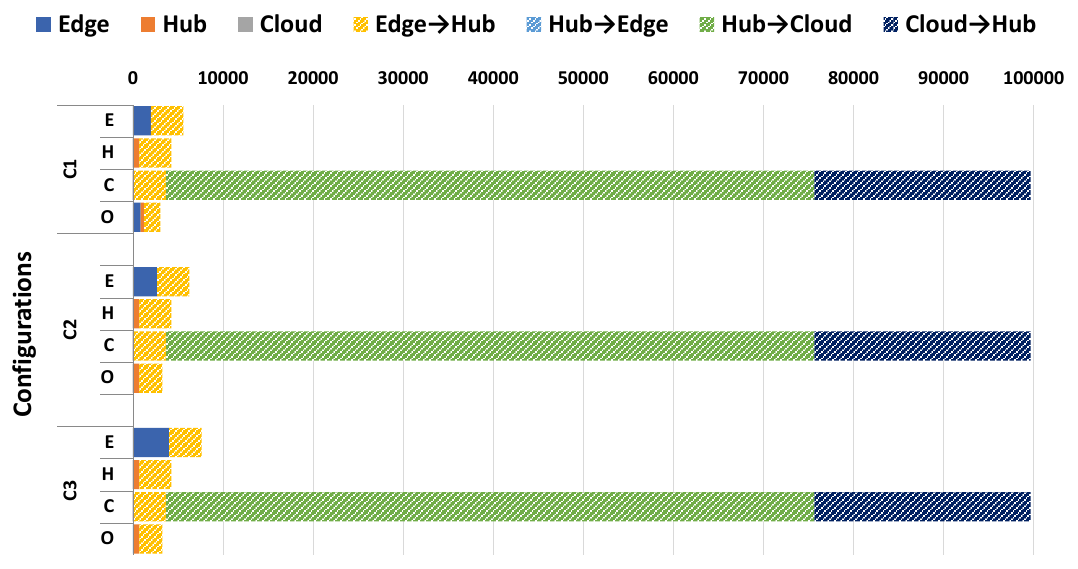}%
        \caption{Latency (in ms) per configuration (Run 2)}%
        \label{results3}%
    \end{subfigure}\hfill%
    \begin{subfigure}[b]{.495\textwidth}
        \includegraphics[width=\columnwidth, height=1.8in]{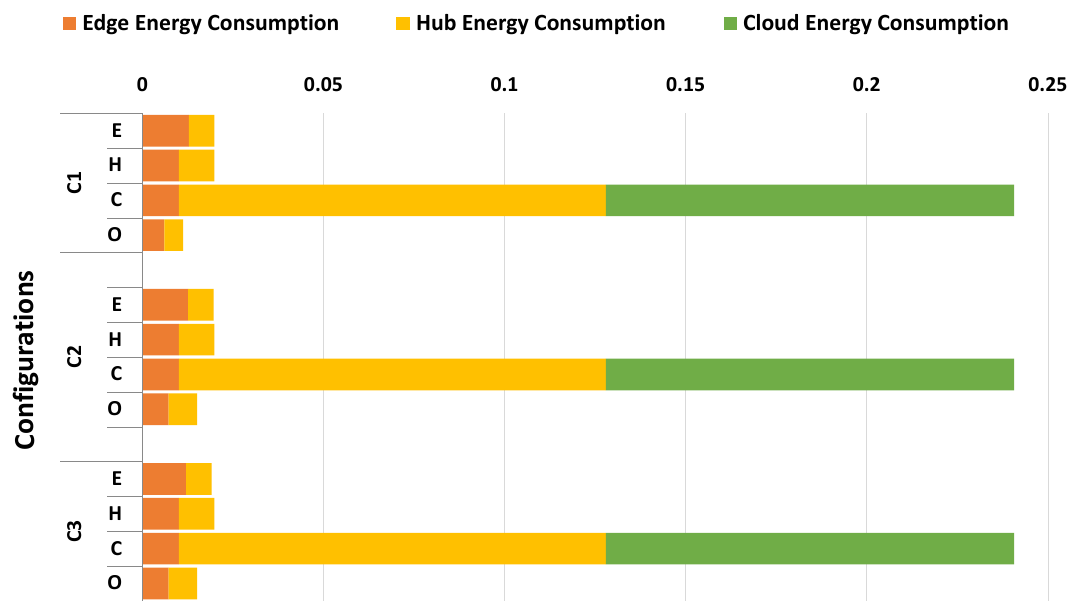}%
        \caption{Energy consumption (in Wh) per configuration (Run 2)}%
        \label{results4}%
    \end{subfigure}\hfill%
    \caption{Real-world use-case scenario: Latency and energy consumption when minimizing overall latency.}
    \label{real_perf}
\end{figure*}

The ETFG of the real-world application was used to formulate and solve the task allocation problem, considering the minimization of either latency or energy.
\cref{real_perf} presents the experimental results with respect to latency and energy consumption,  when the optimization objective was the minimization of overall latency, for all three device configurations (C1, C2, and C3), and for both sets of bandwidth and energy parameters (Run 1 and Run 2). For each configuration, in addition to the optimal task allocation across all three devices (denoted by O), we also examined three extreme task allocation cases where all tasks (except tasks 1 and 15, which required fixed allocation) were allocated on the edge device $\mathrm{e}$, the hub device $\mathrm{h}$, or the cloud server $\mathrm{c}$ (denoted by E, H, and C, respectively). \cref{results1,results3} report the overall latency for Run 1 and Run 2, respectively. They illustrate the computational latency on each device (denoted by Edge, Hub, and Cloud), as well as the communication latency over each communication channel (denoted by $\mathrm{Edge} \rightarrow \mathrm{Hub}$, $\mathrm{Hub} \rightarrow \mathrm{Edge}$, $\mathrm{Hub} \rightarrow \mathrm{Cloud}$, and $\mathrm{Cloud} \rightarrow \mathrm{Hub}$). The computational (communication) latency is depicted in solid (patterned) color. 
\cref{results2,results4} show the overall energy consumption for Run 1 and Run 2, respectively. The total energy consumption of each device is reported as the sum of the utilized computational and communication energy.

\begin{figure*}[t]
    \centering
    \begin{subfigure}{0.95\textwidth}
        \includegraphics[width=\columnwidth]{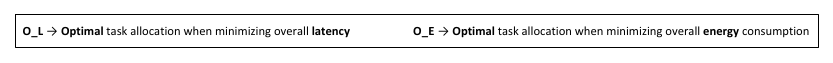}%
    \end{subfigure}\hfill%
    \begin{subfigure}[b]{.495\textwidth}
        \includegraphics[width=\columnwidth, height=1.8in]{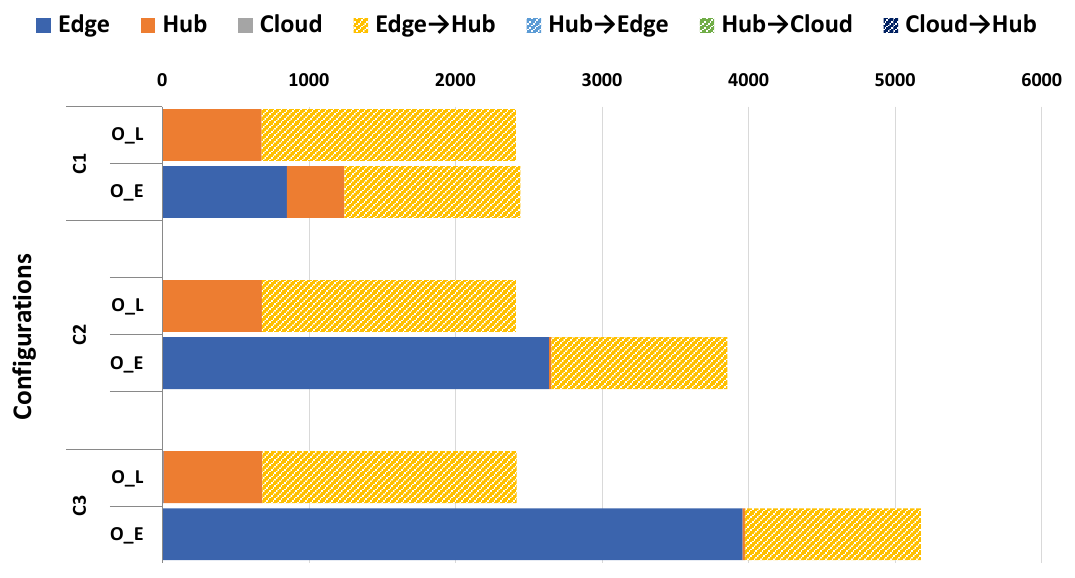}%
        \caption{Latency (in ms) per configuration (Run 1)}%
        \label{results5}%
    \end{subfigure}\hfill%
    \begin{subfigure}[b]{.495\textwidth}
        \includegraphics[width=\columnwidth, height=1.8in]{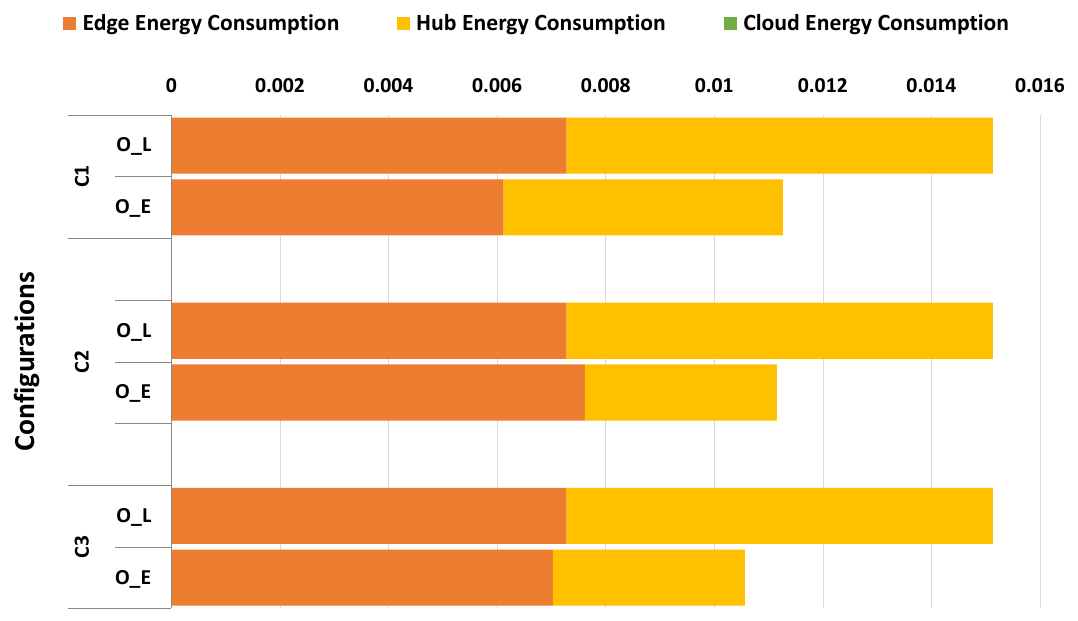}%
        \caption{Energy consumption (in Wh) per configuration (Run 1)}%
        \label{results6}%
    \end{subfigure}\hfill 
    \begin{subfigure}[b]{.495\textwidth}
        \includegraphics[width=\columnwidth, height=1.8in]{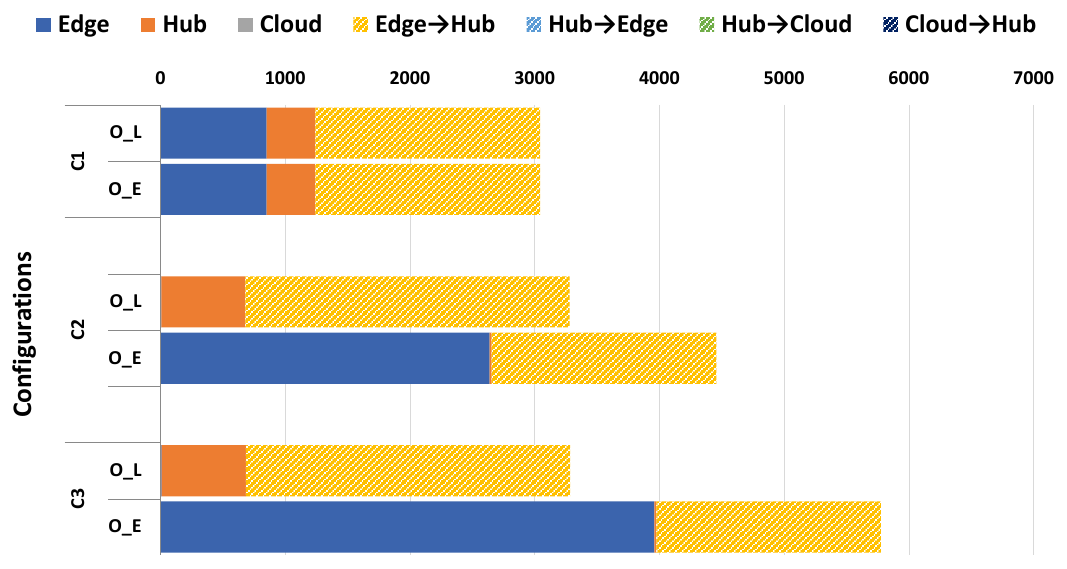}%
        \caption{Latency (in ms) per configuration (Run 2)}%
        \label{results7}%
    \end{subfigure}\hfill%
    \begin{subfigure}[b]{.495\textwidth}
        \includegraphics[width=\columnwidth, height=1.8in]{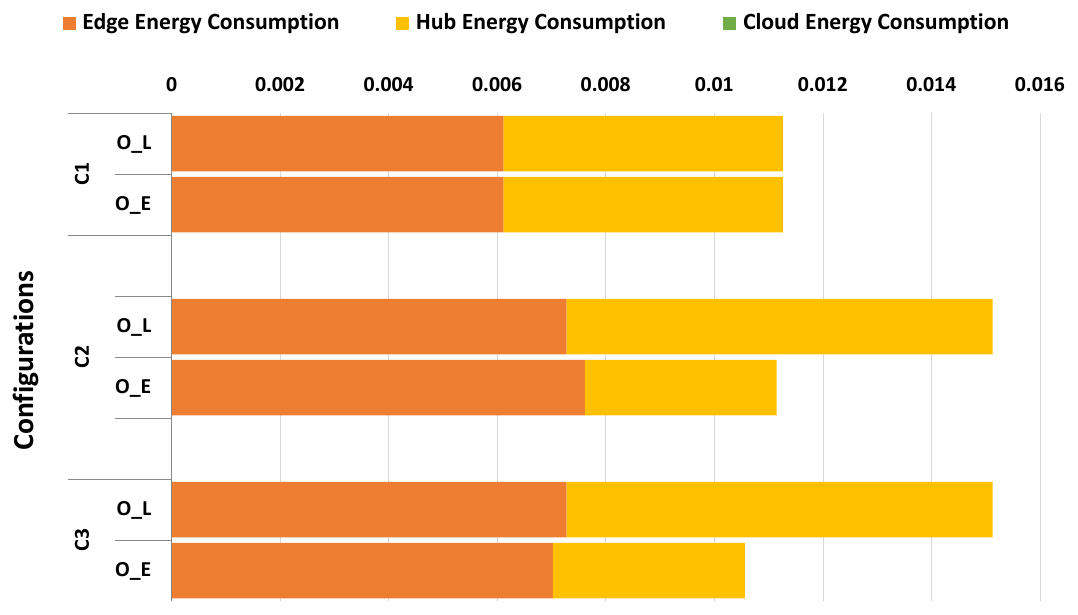}%
        \caption{Energy consumption (in Wh) per configuration (Run 2)}%
        \label{results8}%
    \end{subfigure}\hfill%
   \caption{Real-world use-case scenario: Comparison of latency and energy consumption between cases where the optimization objective was the minimization of either latency (O\_L) or energy (O\_E).}
    \label{real_energy}
\end{figure*}


The results in \cref{real_perf} show that the optimal task allocation (case O) returned by our framework, yielded not only the minimum overall latency, which was the optimization objective, but also the lowest overall energy consumption, for all configurations and runs. 
With respect to latency, \cref{results1} (Run 1) reveals that the second-best performance was observed in case H. On the other hand, the latency in case C was comparatively high and approximately the same for all configurations, due to the fixed allocation of tasks 1 and 15 on devices $\mathrm{e}$ and $\mathrm{h}$, respectively. This resulted in increased latency for communication channels $\mathrm{Edge} \rightarrow \mathrm{Hub}$, $\mathrm{Hub} \rightarrow \mathrm{Cloud}$, and $\mathrm{Cloud} \rightarrow \mathrm{Hub}$, as the remaining tasks were allocated on device $\mathrm{c}$. Despite this, case C provided the worst performance only for configuration C1. The worst performance for C2 and C3 was observed in case E. The reason was that Odroid XU4 and  Raspberry Pi 3 Model B used as edge devices in configurations C2 and C3, respectively, are slower than Jetson TX2 used in C1, thus leading to a higher overall latency.
In \cref{results3} (Run 2), the second-best performance was again observed in case H. However, in contrast to Run 1, the worst performance was always observed in case C, due to the lower bandwidth of communication channels $\mathrm{Hub} \rightarrow \mathrm{Cloud}$ and $\mathrm{Cloud} \rightarrow \mathrm{Hub}$.
With respect to energy, \cref{results2,results4} show that the highest energy consumption was observed in case C. This was due to the significantly higher communication demands, compared to the other cases.


\cref{real_energy} illustrates the comparison of latency and energy consumption between cases where the optimization objective was the minimization of either latency (O\_L) or energy (O\_E), for all configurations (C1, C2, and C3) and runs (Run 1 and Run 2).
The results in \cref{results5,results6,results7,results8} are reported in the same manner as in \cref{real_perf}. They demonstrate how each optimization objective affected the overall latency and energy.
For example, \cref{results5,results7} show a trend in the case of O\_E to allocate as many tasks as possible on device $\mathrm{e}$, in an attempt to minimize inter-task communication and thus the utilized energy.
Furthermore, it can be observed in \cref{results5} that for configuration C1, Run 1, the overall latency was approximately the same for both optimization objectives, O\_L and O\_E, even though the optimal task allocation was different in each case. However, due to the different allocation of the tasks, the resulting energy consumption differed between the two cases, as shown in \cref{results6}. On the other hand, for configuration C1, Run 2, both optimization objectives yielded the same optimal task allocation. Hence, the resulting overall latency and energy were the same in both cases, as showcased in \cref{results7,results8}, respectively.

\begin{table}[t]
\centering
\caption{Solver execution time for real-world use-case scenario.}
\resizebox{\columnwidth}{!}{
\footnotesize
\begin{tabular}{@{\extracolsep{4pt}}crrrr@{}} 
\toprule

\multirow{3}{*}{Configuration} & \multicolumn{4}{c}{Optimization Objective} \\
\cline{2-5}
& \multicolumn{2}{c}{Latency} & \multicolumn{2}{c}{Energy Consumption} \\
\cline{2-3} \cline{4-5}

 & \multicolumn{1}{c}{Run 1} & \multicolumn{1}{c}{Run 2} & \multicolumn{1}{c}{Run 1} & \multicolumn{1}{c}{Run 2} \\

\hline


C1	& 7.39\,ms	& 11.31\,ms	& 10.69\,ms	& 13.43\,ms \\
C2	& 8.10\,ms	& 13.86\,ms	& 11.27\,ms	& 12.42\,ms \\
C3	& 7.48\,ms	& 13.86\,ms	& 10.90\,ms	& 12.41\,ms \\

\bottomrule
\end{tabular}
}
\label{table:realTimes}
\end{table}

\begin{figure*}[t]
    \centering
    \includegraphics[width=.85\textwidth]{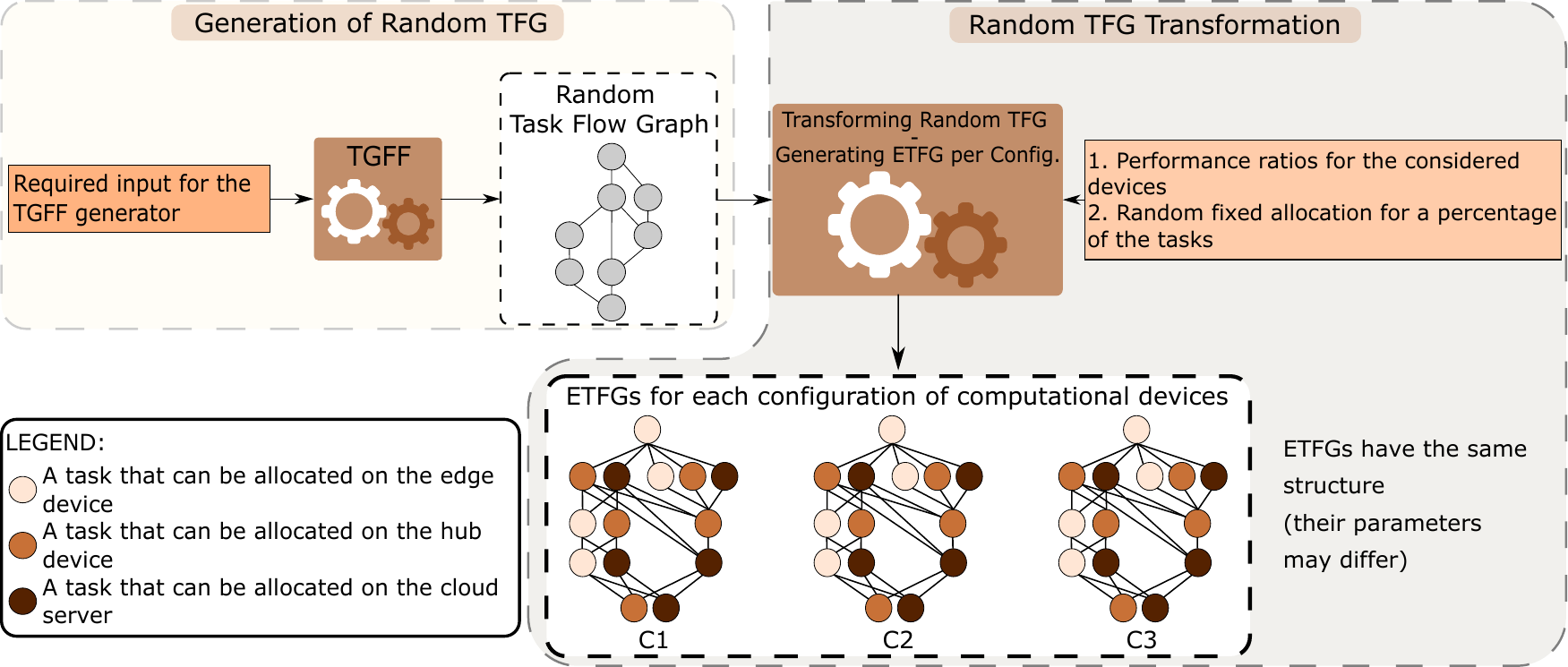}
    \caption{Overview of random TFG generation and transformation. The generation of random TFGs is presented in \cref{subsubsec:randomTFGs}. The transformation of random TFGs into ETFGs is described in \cref{subsubsec:syntheticParams}.}
    \label{overview_tgf_generation}
\end{figure*}

The above observations indicate that a straightforward derivation of a suitable task allocation cannot be based on the optimization objective alone, as other factors play a decisive role, such as the communication and fixed allocation requirements of the tasks, as well as the device and communication channel characteristics. Therefore, the use of the proposed application-driven optimization framework is beneficial in that regard. In addition to providing the optimal task allocation for the minimization of either latency or energy, it also enables design space exploration with respect to different device configurations and their connectivity.
As reported by the Gurobi solver, our framework required 152 variables for the examined real-world application. It required 149 constraints when minimizing the overall latency, and 150 constraints when minimizing the overall energy consumption. In both cases, it required an average time of 11.09\,ms in order to return a solution. Specifically, the solver execution time ranged between 7.39\,ms and 13.86\,ms (as shown in \cref{table:realTimes}), which is short and practical, as this is a design-time (i.e., offline) approach.


\subsection{Synthetic benchmarks}\label{pseudo-random}
We developed various synthetic benchmarks (i.e., synthetic ETFGs) with different structures, sizes, and parameters, suitable for the targeted  edge/hub/cloud environment. Since the applications motivating this work typically use machine learning inference and/or time-bound stream processing methods, we considered the following relevant structure types: parallel (denoted by P), serial (denoted by S), and a mixture of both (denoted by M). 
The initial TFGs were generated randomly (\cref{subsubsec:randomTFGs}).
To transform a randomly generated TFG into the corresponding ETFGs (one ETFG per device configuration), we extended our TFG transformation technique (\cref{subsubsec:syntheticParams}).
The parameters of the resulting ETFGs were determined using workload information intended for such applications, with the aid of relevant device benchmarks.
An outline of the process for generating a random TFG and subsequently transforming it into ETFGs, is illustrated in \cref{overview_tgf_generation}.

\subsubsection{Generation of random TFGs}\label{subsubsec:randomTFGs}
\begin{figure}[t]
    \centering
    \begin{subfigure}{.33\columnwidth}
        \centering
        \includegraphics[height=1in]{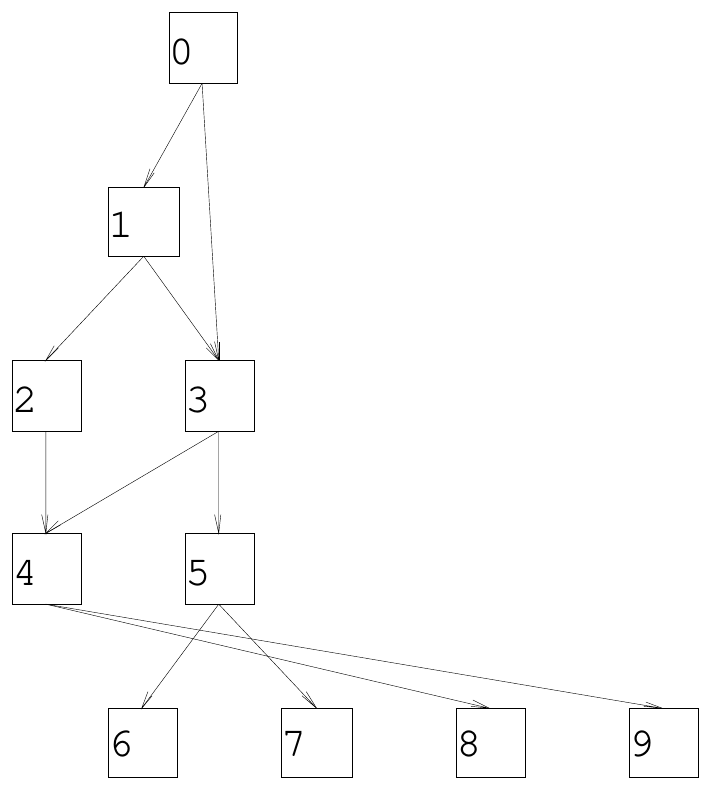}
        \caption{P1.1}
        \label{randomTaskGraphsa}
    \end{subfigure}\hfill%
    \begin{subfigure}{.33\columnwidth}
        \centering
        \includegraphics[height=2in]{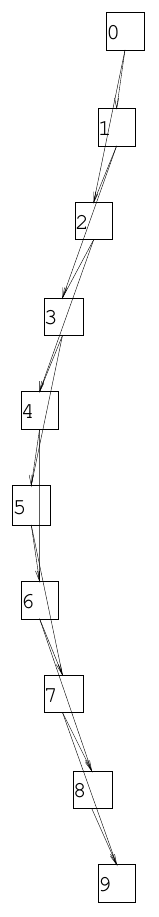}
        \caption{S1.1}
        \label{randomTaskGraphsb}
    \end{subfigure}\hfill%
    \begin{subfigure}{.33\columnwidth}
        \centering
        \includegraphics[height=2.22in]{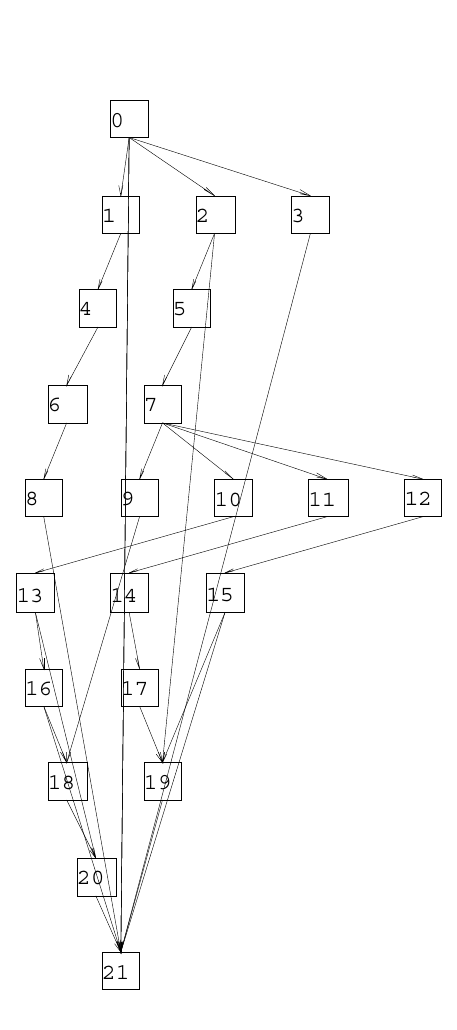}
        \caption{M1.1}
        \label{randomTaskGraphsc}
    \end{subfigure}\hfill%
    \caption{Randomly generated TFGs with (a) parallel, (b) serial, and (c) mixed structure type, as described in \cref{table:tfg_properties}.}
    \label{randomTaskGraphs}
\end{figure}

We used the Task Graphs for Free (TGFF) generator \cite{tgff, tgff2} to randomly generate the initial TFGs. The input required by the TGFF generator included the number of nodes (tasks) and the node maximum in/out degree (i.e., the maximum number of incoming/outgoing arcs per node), as shown in the second and third columns of \cref{table:tfg_properties}, respectively. The particular input was used to generate a variety of representative TFGs \cite{synthetic_taskgraphs}.
Specifically, we generated six random TFGs for each structure type, with 10, 100, and 1000 nodes. For each number of nodes, two TFG variants were generated, using different maximum in/out degrees. 
The identifier of each randomly generated TFG (shown in the first column of \cref{table:tfg_properties}) denotes the structure type (P, S, or M), the number of nodes (order of magnitude), and the variant of the TFG. 
For example, P2.1 is the first variant of a TFG with parallel structure and 100 nodes.
Due to the nature of the targeted applications, after the generation of each TFG, we set a percentage of its tasks to require fixed allocation on devices $\mathrm{e}$ and $\mathrm{h}$, as shown in the seventh and eighth columns of \cref{table:tfg_properties}, respectively. In each case of fixed allocation, the tasks were selected randomly.
Graphical representations of P1.1, S1.1, and M1.1 TFGs are depicted in \cref{randomTaskGraphs}.

\begin{table*}[t]
\centering
\caption{Randomly generated TFGs and their corresponding ETFGs (Synthetic Benchmarks).}
\label{table:tfg_properties}
\footnotesize
\resizebox{0.85\textwidth}{!}{
\begin{tabular}{@{\extracolsep{4pt}}ccccccccccc@{}}
\toprule
 \multicolumn{3}{c}{Input for TGFF} & \multicolumn{3}{c}{Generated} & \multicolumn{2}{c}{Post} & \multicolumn{3}{c}{Corresponding} \\
 \multicolumn{3}{c}{Generator} & \multicolumn{3}{c}{TFG} & \multicolumn{2}{c}{Generation} & \multicolumn{3}{c}{ETFG} \\
\cline{1-3} \cline{4-6} \cline{7-8} \cline{9-11}
\multirow{2}{*}{TFG\#} & \multirow{2}{*}{\begin{tabular}[c]{@{}c@{}}  \#Nodes \end{tabular}}  & \multirow{2}{*}{\begin{tabular}[c]{@{}c@{}}Max\\In/Out-Degree \end{tabular}} &  \multirow{2}{*}{\begin{tabular}[c]{@{}c@{}} \#Nodes/Arcs \end{tabular}} & \multirow{2}{*}{\begin{tabular}[c]{@{}c@{}}Avg.\\In/Out-Degree \end{tabular}}  & \multirow{2}{*}{\begin{tabular}[c]{@{}c@{}}Depth/\\Max Width \end{tabular}} & \multicolumn{2}{c}{\begin{tabular}[c]{@{}c@{}}Fixed\\Allocation (\%) \end{tabular}} & \multirow{2}{*}{\begin{tabular}[c]{@{}c@{}} \#Nodes/Arcs \end{tabular}} & \multirow{2}{*}{\begin{tabular}[c]{@{}c@{}} \#Variables/\\ Constraints \end{tabular}} & \multirow{2}{*}{\begin{tabular}[c]{@{}c@{}} Avg. Exec.\\Time (s) \end{tabular}}\\
 &  &  &  &  &  & $\mathrm{e}$ & $\mathrm{h}$ & & &\\

\hline


P1.1 & 10 		& 2\,/\,2 		& 10\,/\,11 			& 1.20\,/\,1.20 	& 5\,/\,4 		& 5 & 2 	& 28\,/\,93 			& 121\,/\,118           & 0.01      \\	
P1.2 & 10 		& 2\,/\,2 		& 9\,/\,10 				& 1.22\,/\,1.22 	& 2\,/\,4 		& 4 & 2 	& 27\,/\,90 			& 117\,/\,112           & 0.01      \\  
P2.1 & 100 		& 2\,/\,3 		& 100\,/\,129 			& 1.30\,/\,1.30 	& 13\,/\,11 	& 4 & 2 	& 288\,/\,1077 			& 1365\,/\,1261         & 0.03      \\  
P2.2 & 100 		& 5\,/\,3 		& 99\,/\,136 			& 1.38\,/\,1.38 	& 6\,/\,24 		& 3 & 1 	& 289\,/\,1170 			& 1459\,/\,1335         & 0.03      \\	
P3.1 & 1000 	& 2\,/\,3 		& 999\,/\,1232 			& 1.23\,/\,1.23 	& 20\,/\,89 	& 5 & 2 	& 2857\,/\,10078 		& 12935\,/\,11812       & 0.25      \\	
P3.2 & 1000 	& 5\,/\,6 		& 1001\,/\,1553 		& 1.55\,/\,1.55 	& 20\,/\,145 	& 4 & 2 	& 2889\,/\,12913 		& 15802\,/\,14553       & 0.44      \\  
	 &  		&  		&  						&  				&  			&   &   &  		&	          \\                    
S1.1 & 10 		& 2\,/\,2 		& 10\,/\,17 			& 1.80\,/\,1.80 	& 10\,/\,1 		& 4 & 2 	& 30\,/\,153 			& 183\,/\,181           & 0.01      \\	
S1.2 & 10 		& 5\,/\,5 		& 11\,/\,40 			& 3.72\,/\,3.72 	& 11\,/\,1 		& 4 & 3 	& 33\,/\,360 			& 393\,/\,390           & 0.01      \\	
S2.1 & 100 		& 2\,/\,2 		& 100\,/\,197 			& 1.98\,/\,1.98 	& 100\,/\,1 	& 4 & 3 	& 286\,/\,1605 			& 1891\,/\,1813         & 0.04      \\	
S2.2 & 100 		& 5\,/\,5 		& 101\,/\,490 			& 4.86\,/\,4.86 	& 101\,/\,1 	& 1 & 3 	& 295\,/\,4170 			& 4465\,/\,4380         & 0.10      \\	
S3.1 & 1000 	& 2\,/\,2 		& 1000\,/\,1997 		& 1.99\,/\,1.99 	& 1000\,/\,1 	& 5 & 3 	& 2840\,/\,16089 		& 18929\,/\,18097       & 0.40      \\	
S3.2 & 1000 	& 5\,/\,5 		& 998\,/\,4975 			& 4.98\,/\,4.98 	& 998\,/\,1 	& 2 & 2 	& 2914\,/\,42415 		& 45329\,/\,44419       & 1.80      \\	
	 &  		&  				&  					&  				&  			&   &   &  	&             \\	                
M1.1 & 10 		& 9\,/\,4 		& 22\,/\,33 			& 1.54\,/\,1.54 	& 6\,/\,5 		& 4 & 1 	& 64\,/\,261 			& 325\,/\,313           & 0.02      \\   
M1.2 & 10 		& 8\,/\,2 		& 55\,/\,65 			& 1.20\,/\,1.20 	& 14\,/\,6 		& 1 & 2 	& 161\,/\,561 			& 722\,/\,679           & 0.02      \\   
M2.1 & 100 		& 10\,/\,3 		& 109\,/\,141 			& 1.30\,/\,1.30 	& 8\,/\,10 		& 1 & 3 	& 319\,/\,1210 			& 1529\,/\,1436         & 0.03      \\   
M2.2 & 100 		& 8\,/\,5 		& 122\,/\,147 			& 1.21\,/\,1.21 	& 7\,/\,15 		& 2 & 2 	& 358\,/\,1252 			& 1610\,/\,1504         & 0.03      \\   
M3.1 & 1000 	& 12\,/\,4 		& 1000\,/\,1224 		& 1.23\,/\,1.23 	& 31\,/\,40 	& 4.80 & 2 	& 2864\,/\,10055 		& 12919\,/\,12063       & 0.24      \\	 
M3.2 & 1000 	& 18\,/\,5 		& 1017\,/\,1181 		& 1.16\,/\,1.16 	& 29\,/\,56 	& 1.86 & 1 	& 2993\,/\,10218 		& 13211\,/\,12260       & 0.24      \\	 

\bottomrule
\end{tabular}
}
\end{table*}

\begin{table*}[t]
\centering
\caption{Performance scores and performance ratios of computational devices.}
\label{table:benchmarks}
\footnotesize
\begin{tabular}{c|r|rr|rr|rr|rr} 
\toprule
 & \multicolumn{9}{c}{Device $k$} \\
\hline
$\mathrm{e}$/$\mathrm{h}$/$\mathrm{c}$ & \multicolumn{1}{c|}{$\mathrm{e}$} & \multicolumn{2}{c|}{$\mathrm{e}$} & \multicolumn{2}{c|}{$\mathrm{e}$} & \multicolumn{2}{c|}{$\mathrm{h}$} & \multicolumn{2}{c}{$\mathrm{c}$}\\

\hline 

Rank                        & \multicolumn{1}{c|}{1}                 & \multicolumn{2}{c|}{2}                 & \multicolumn{2}{c|}{3}                 & \multicolumn{2}{c|}{4}                     & \multicolumn{2}{c}{5}\\

\hline 

\multirow{2}{*}{Device}            &   \multicolumn{1}{c|}{Raspberry Pi 3*}	&   \multicolumn{2}{c|}{\multirow{2}{*}{Odroid XU4}}  				    &   \multicolumn{2}{c|}{\multirow{2}{*}{Jetson TX2}}       		&   \multicolumn{2}{c|}{\multirow{2}{*}{Mi Notebook Pro}}				    &  \multicolumn{2}{c}{HPE ProLiant}\\ 
\multirow{2}{*}{Benchmark} &   \multicolumn{1}{c|}{Model B} & &  & & & & & \multicolumn{2}{c}{DL580 Gen10}		\\
\cline{2-10}

 & \multicolumn{1}{c|}{Score} & \multicolumn{1}{c}{Score} & \multicolumn{1}{c|}{$\phi_{k}$}  & \multicolumn{1}{c}{Score} & \multicolumn{1}{c|}{$\phi_{k}$}  & \multicolumn{1}{c}{Score} & \multicolumn{1}{c|}{$\phi_{k}$}  & \multicolumn{1}{c}{Score} & \multicolumn{1}{c}{$\phi_{k}$}\\
 
\hline

R Benchmark                          & 3.80\,s 	\hspace{10pt}	 & 2.16\,s    	&	1.76		& 2.22\,s    	&	1.71		& 0.50\,s    		&	7.60		&   0.26\,s     &	14.62	\\
Scikit-Learn (Phoronix)              & 1373.54\,s \hspace{10pt}	 & 629.74\,s 	&	2.18		& 268.09\,s 	&	5.12		& 10.11\,s 			&	135.86	&   7.64\,s  	&	179.78	\\
TensorFlow (Phoronix)                & 12648.38\,s \hspace{10pt} & 1775.29\,s    &	7.12		& 781.46\,s 	&	16.19		& 169.70\,s 		&	74.53	&   59.82\,s    &	211.44	\\
NumPy (Phoronix)                     & 10.63\,pts \hspace{10pt}  & 24.93\,pts    &	2.35		& 52.81\,pts    &	4.97		& 284.68\,pts   	&	26.78	&   405.71\,pts &	38.17	\\
Geekbench 5                          & 398.00\,pts \hspace{10pt} & 626.00\,pts  	&	1.57		& \multicolumn{1}{c}{-} 			&	\multicolumn{1}{c|}{-} 			& 3612.00\,pts 		& 	9.08	&   33323.00\,pts  &	83.73	\\

\hline
Avg. Performance Ratio  & & \multicolumn{2}{r|}{$\phi_{k}^{\mathrm{avg}} = 2.99$} & \multicolumn{2}{r|}{$\phi_{k}^{\mathrm{avg}} = 6.99$} & \multicolumn{2}{r|}{$\phi_{k}^{\mathrm{avg}} = 50.77$}  & \multicolumn{2}{r}{$\phi_{k}^{\mathrm{avg}} = 105.55$}\\

\bottomrule

\multicolumn{6}{l}{\scriptsize{*Used as reference device $\hat{\mathrm{e}}$.}}\\
\end{tabular}
\end{table*}

\subsubsection{Transformation of random TFGs into ETFGs}\label{subsubsec:syntheticParams}
We transformed each random TFG by extending our approach to generate an ETFG for each device configuration (C1, C2, and C3). 
For a specific TFG, the resulting ETFGs had the same structure, but their parameters could differ.
Specifically, to randomly assign appropriate values to the ETFG candidate node and arc parameters for each configuration, we first ran on each device the benchmarks shown in \cref{table:benchmarks}.
The Phoronix Test Suite (Scikit-Learn, TensorFlow, and NumPy) and R Benchmark \cite{phoronix, openbenchmarking} were run on all devices in each configuration.
Geekbench 5 was run on devices $\mathrm{h}$ and $\mathrm{c}$, whereas for all variants of device $\mathrm{e}$, we used the scores reported in \cite{geekbench5} (except for Jetson TX2, for which a score is not available). 
The NumPy and Geekbench 5 results are shown in points (pts), where a higher score indicates better performance (in terms of latency). The remaining benchmark results are reported in seconds (s), where a lower score represents better performance. 
We ranked the devices based on their performance scores, from the slowest (rank 1) to the fastest (rank 5), as shown in \cref{table:benchmarks}. We used the slowest device, Raspberry Pi 3 Model B, as a reference device (denoted by $\hat{\mathrm{e}}$) to determine the performance ratio $\phi_{k}$ of the other devices. For the benchmark results in seconds, $\phi_{k}$ is the ratio of the score of device $\hat{\mathrm{e}}$ to the score of device $k$. For the results in points, $\phi_{k}$ is the ratio of the score of device $k$ to the score of device $\hat{\mathrm{e}}$.

Subsequently, we determined the computational latency $L_{ik}$ and power consumption $P_{ik}$ for each candidate node $N_{ik}$. We first assigned a random value to $L_{i \hat{\mathrm{e}}}$ and $P_{i \hat{\mathrm{e}}}$, selected from the respective measurements for device $\hat{\mathrm{e}}$ in the real-world use-case scenario. 
For the remaining devices, we set $L_{ik} = L_{i \hat{\mathrm{e}}} / \phi_{k}^{\mathrm{avg}}$ and $P_{ik} = P_{i \hat{\mathrm{e}}} \, \phi_{k}^{\mathrm{avg}}$, where $\phi_{k}^{\mathrm{avg}}$ is the average performance ratio of device $k$ across all device benchmarks, as shown in \cref{table:benchmarks}.
In each case, we ensured that $P_{ik}$ was in the interval $( P_{k}^{\mathrm{idle}}, P_{k}^{\mathrm{max}} ]$, by adjusting it accordingly. Specifically, if $P_{ik} \leq P_{k}^{\mathrm{idle}}$ or $P_{ik} > P_{k}^{\mathrm{max}}$, we adjusted it to $P_{k}^{\mathrm{idle}} (1 + \alpha)$ and $P_{k}^{\mathrm{max}} (1 - \alpha)$, respectively. The adjustment factor $\alpha$ was randomly selected in the range $[ 0.1\%, 0.5\% ]$.
$P_{k}^{\mathrm{idle}}$ and $P_{k}^{\mathrm{max}}$ were defined through profiling and represent the idle and maximum power consumption of device $k$, respectively. 
The candidate node parameters $M_{i}$, $S_{i}$, and $D_{i}$ were selected randomly from the corresponding measurements in the real-world scenario.
On the other hand, the remaining candidate node and arc parameters, $E_{ik}$, $NC_{i}$, $\delta_{ik \rightarrow jl}^{m}$, $CL_{ik \rightarrow jl}$, and $CE_{ik \rightarrow jl}$ were determined as explained in \cref{real_app}. For $CL_{ik \rightarrow jl}$ and $CE_{ik \rightarrow jl}$ we used the bandwidth and energy parameters of Run 1 in \cref{table:energy_table}. The number of nodes and arcs in each ETFG are shown in the ninth column of \cref{table:tfg_properties}.

\subsubsection{Framework evaluation -- Synthetic benchmarks}\label{pseudo-random-results}

\begin{figure*}[ht]
    \centering
    \begin{subfigure}[b]{.495\textwidth}
        \includegraphics[width=\columnwidth, height=1.8in]{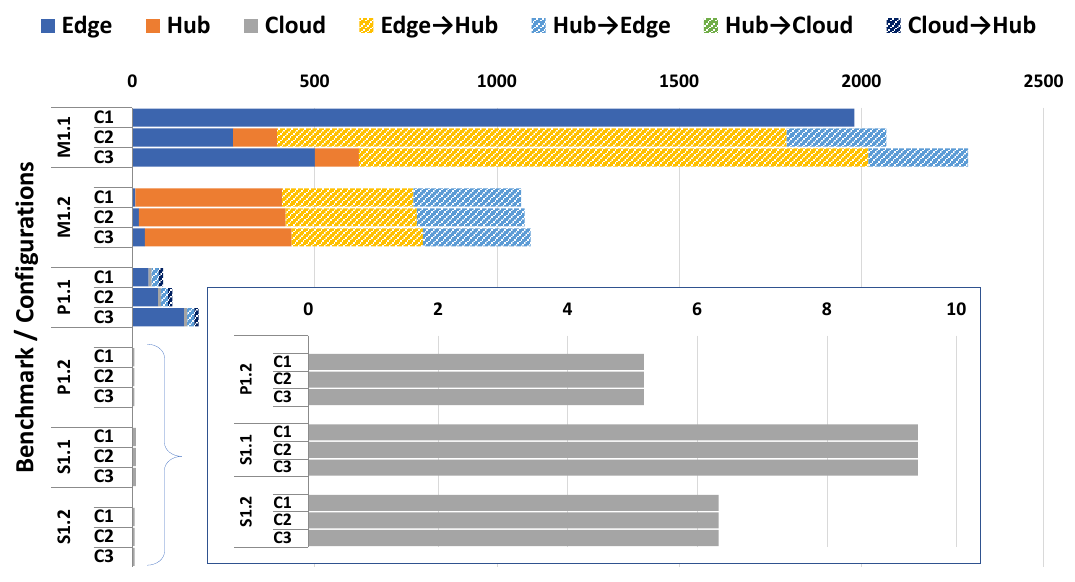}%
        \caption{Latency (in\,ms) per benchmark (10 nodes) and configuration}%
        \label{results9}%
    \end{subfigure}\hfill%
    \begin{subfigure}[b]{.495\textwidth}
        \includegraphics[width=\columnwidth, height=1.8in]{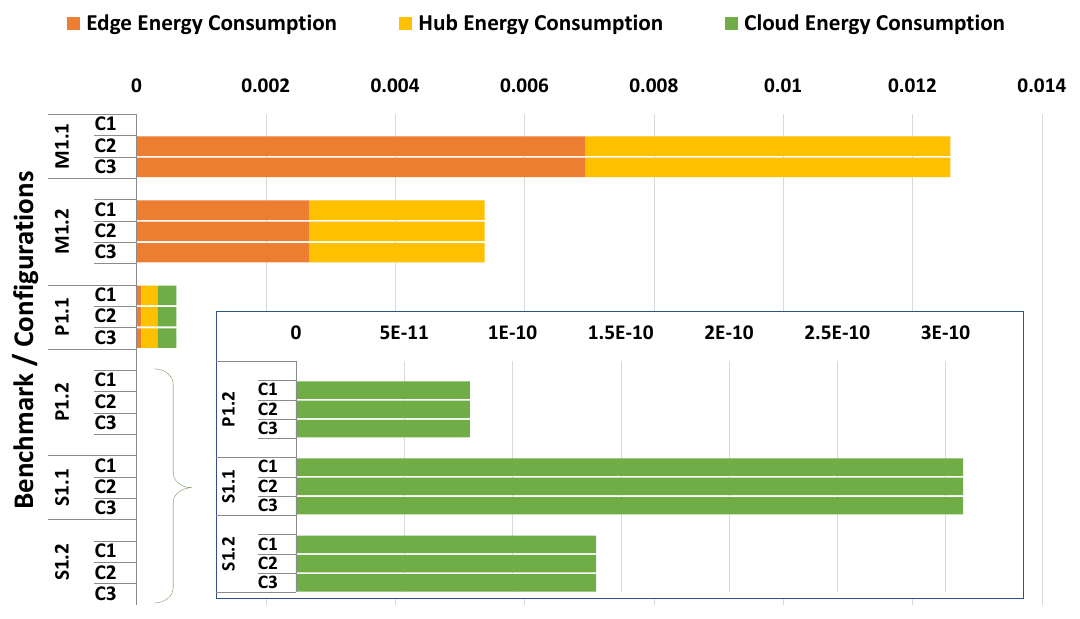}%
        \caption{Energy consumption (in\,Wh) per benchmark (10 nodes) and configuration}%
        \label{results10}%
    \end{subfigure}\hfill 
    \begin{subfigure}[b]{.495\textwidth}
        \includegraphics[width=\columnwidth, height=1.8in]{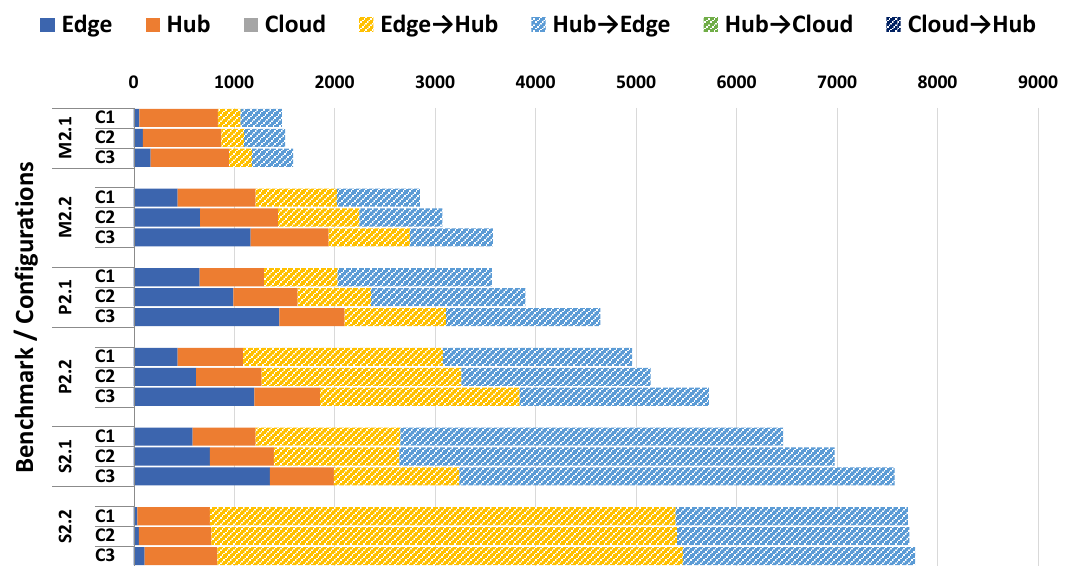}%
        \caption{Latency (in\,ms) per benchmark (100 nodes) and configuration}%
        \label{results11}%
    \end{subfigure}\hfill%
    \begin{subfigure}[b]{.495\textwidth}
        \includegraphics[width=\columnwidth, height=1.8in]{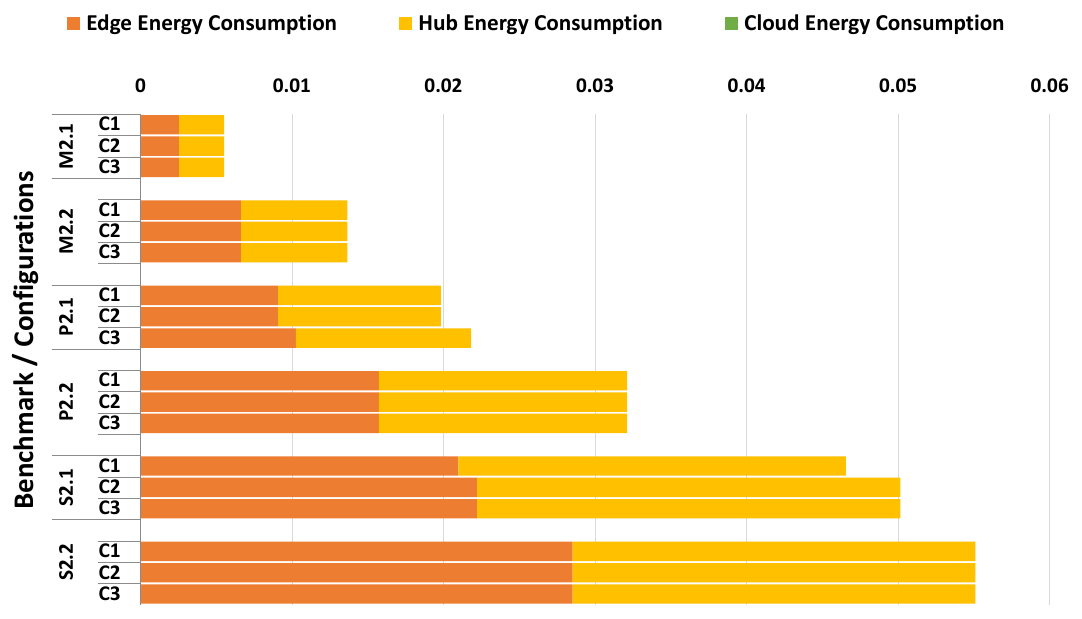}%
        \caption{Energy consumption (in\,Wh) per benchmark (100 nodes) and configuration}%
        \label{results12}%
    \end{subfigure}\hfill%
    \begin{subfigure}[b]{.495\textwidth}
        \includegraphics[width=\columnwidth, height=1.8in]{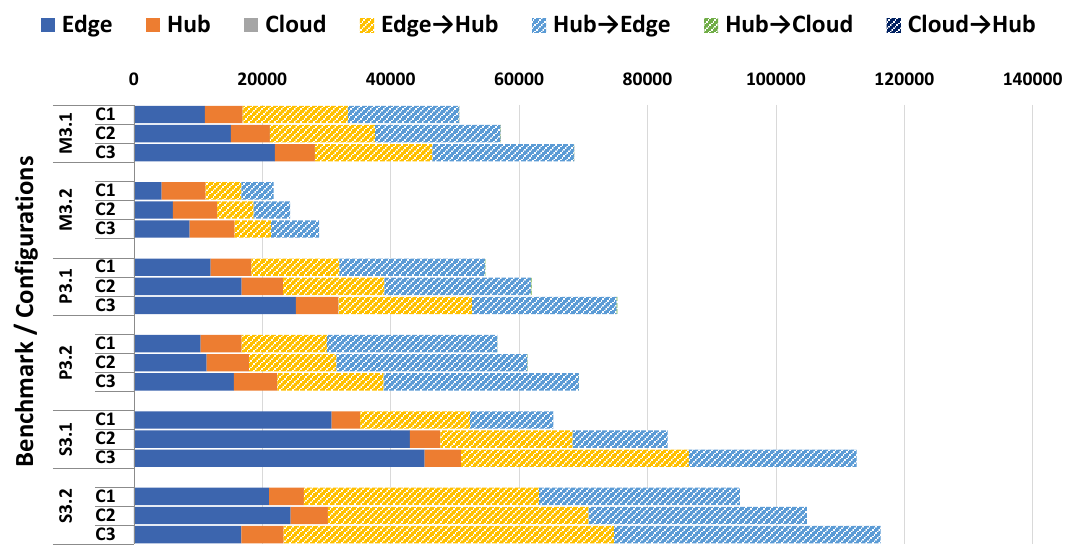}%
        \caption{Latency (in\,ms) per benchmark (1000 nodes) and configuration}%
        \label{results13}%
    \end{subfigure}\hfill%
    \begin{subfigure}[b]{.495\textwidth}
        \includegraphics[width=\columnwidth, height=1.8in]{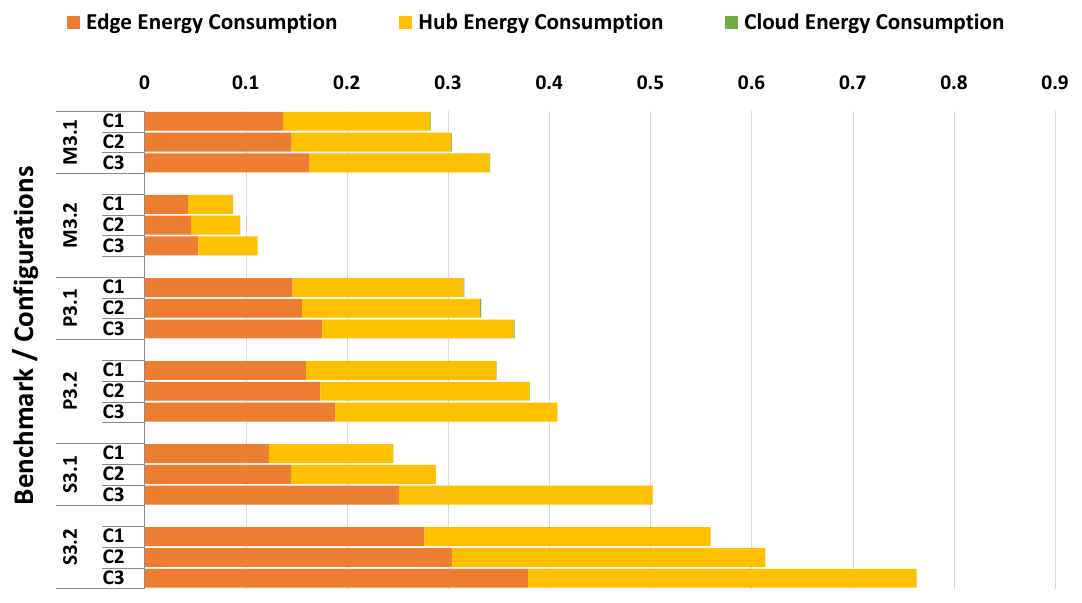}%
        \caption{Energy consumption (in\,Wh) per benchmark (1000 nodes) and configuration}%
        \label{results14}%
    \end{subfigure}\hfill 
    \caption{Synthetic benchmarks:
    Latency and energy consumption when minimizing overall latency.
    }
    \label{benchmarks}
    \vspace{-10pt}
\end{figure*}

The resulting ETFGs, denoted by the identifier of their corresponding TFG, were used as synthetic benchmarks. 
\cref{benchmarks} presents the evaluation results with respect to latency and energy consumption when minimizing overall latency, for all three configurations (C1, C2, and C3), using the bandwidth and energy parameters of Run 1. The results are presented in the same fashion as in the real-world use-case (\cref{case_study-results}).
It can be observed that the overall latency and energy consumption increased as the size of the benchmarks expanded. 
Moreover, \cref{results9} shows that for benchmark M1.1, configuration C1, all tasks were allocated on device $\mathrm{e}$. This led to a significantly lower energy consumption than C2 and C3 for the particular benchmark, as shown in \cref{results10}.
This was due to the utilization of Jetson TX2 as device $\mathrm{e}$ in C1, instead of the slower Odroid XU4 or Raspberry Pi 3 Model B used in C2 and C3, respectively. Therefore, only device $\mathrm{e}$ was required to provide the minimum latency for benchmark M1.1, configuration C1. On the other hand, in configurations C2 and C3 for the specific benchmark, device $\mathrm{h}$ was also utilized in addition to device $\mathrm{e}$, resulting in higher energy consumption than C1, due to the communication between the two devices.

Furthermore, \cref{results9,results10} show that the overall latency and energy for benchmarks M1.1 and M1.2 were significantly higher than those for P1.1, P1.2, S1.1, and S1.2. 
This was due to the considerably greater number of nodes and arcs in M1.1 and M1.2 TFGs compared to the other structure types, as generated by the TGFF generator for the same input (i.e., 10 nodes). 
This led to increased computational and communication requirements for M1.1 and M1.2.
On the other hand, \crefrange{results11}{results14} reveal that for TFGs with 100 and 1000 nodes, serial benchmarks generally did not perform well with respect to latency and energy. 
The reason was that they featured more arcs compared to the respective parallel and mixed benchmarks. Hence, they had comparatively higher communication demands.

Consequently, the structure of a TFG plays a  pivotal role in the examined problem. TFGs with a similar number of nodes but varied structure resulted in diverse task allocations, yielding  different results.
Moreover, inter-task communication was often the dominant factor in the resulting overall latency and energy consumption. These observations are in line with the ones made for the real-world application (\cref{case_study-results}).
The number of variables and constraints, as well as the average time across all configurations required by the Gurobi solver to return a solution for each benchmark are shown in \cref{table:tfg_properties}. The average time ranged between 0.01\,s (benchmarks P1.1, P1.2, S1.1, and S1.2) and 1.8\,s (benchmark S3.2). As this is a design-time framework where task allocation is performed offline, and also considering the NP-hardness of the problem, the solver execution time is short and thus practical. 
Overall, the evaluation of the synthetic benchmarks demonstrated the efficiency and scalability of the proposed approach to applications of different structures and sizes. 
It consistently provided the optimal solution in a short time frame (1.8\,s in the worst case), even for large-scale TFGs with 1000 nodes, requiring over 45\,000 variables and over 44\,000 constraints.

\section{Conclusions}\label{conc}
Motivated by the task allocation challenges in the edge/hub/cloud continuum for applications requiring an edge device, a hub device, and a cloud server, we proposed a comprehensive BILP formulation, encapsulated in an application-driven design-time framework, to optimally and efficiently address the task allocation problem in the particular architecture.
Our approach transforms the task flow graph of an application into an extended form to facilitate the formulation of the examined problem. 
It supports two distinct optimization objectives, the minimization of either latency or energy, while incorporating parameters and constraints often ignored in related studies, such as the computational and communication latency and energy required for the processing of the tasks, as well as the memory, storage, and energy limitations of the devices.
We validated our framework using a real-world application under varied device configurations and communication channel characteristics. We further evaluated its scalability to applications of different structures and sizes utilizing appropriate synthetic benchmarks we developed for this purpose.
Through extensive experimentation, we demonstrated that the proposed method not only yields optimal and scalable results, but it also enables efficient design space exploration with respect to different devices and their connectivity.

\vspace{-10pt}
\section*{Data availability}
\vspace{-5pt}
The datasets of our synthetically generated task flow graphs are publicly available (open-access) at https://doi.org/10.5281/zenodo.10654551.

\vspace{-10pt}
\section*{Acknowledgments}
\vspace{-5pt}
This work has been supported by the European Union’s Horizon 2020 research and innovation programme under grant agreement No. 739551 (KIOS CoE) and from the Government of the Republic of Cyprus through the Cyprus Deputy Ministry of Research, Innovation and Digital Policy.

\vspace{-10pt}
\section*{Appendix A. Supplementary data}
\label{sec:appendix}
\vspace{-5pt}
Supplementary material related to this article can be found online at https://doi.org/10.1016/j.future.2024.02.005.

\biboptions{numbers,sort&compress}
\bibliographystyle{elsarticle-num} 
\vspace{-10pt}
\bibliography{references.bib}

\appendix


\section*{Supplementary material}

# An optimization framework for task allocation in the edge/hub/cloud paradigm

Andreas Kouloumpris[a,b,*], Georgios L. Stavrinides[b], Maria K. Michael[a,b], Theocharis Theocharides[a,b]

[a]*Department of Electrical and Computer Engineering, University of Cyprus, 1 Panepistimiou Avenue, Aglantzia, P.O. Box 20537, Nicosia, 1678, Cyprus*
[b]*KIOS Research and Innovation Center of Excellence, University of Cyprus, 1 Panepistimiou Avenue, Aglantzia, P.O. Box 20537, Nicosia, 1678, Cyprus*

## Appendix A

Table A.1: List of notations used in the proposed framework.

| Notation | Definition |
|---|---|
| $\mathcal{T}$ | Set of tasks of the considered application. |
| $\mathcal{U}$ | Set of computational devices in the target system. |
| $\mathcal{F}_i$ | Subset of computational devices where task $i$ can be allocated ($\mathcal{F}_i \subseteq \mathcal{U}$). |
| $i, j$ | Tasks of the considered application ($i, j \in \mathcal{T}$). |
| $k, l, m$ | Computational devices in the target system ($k, l, m \in \mathcal{U}$). |
| $G = (\mathcal{N}, \mathcal{A})$ | Task flow graph (TFG) of the considered application. |
| $\mathcal{N}$ | Set of nodes in the TFG. |
| $\mathcal{A}$ | Set of arcs in the TFG. |
| $N_i$ | Node in the TFG ($N_i \in \mathcal{N}$) corresponding to task $i \in \mathcal{T}$. |
| $A_{i \to j}$ | Arc in the TFG ($A_{i \to j} \in \mathcal{A}$) connecting nodes $N_i$ and $N_j$. |
| $G' = (\mathcal{N}', \mathcal{A}')$ | Extended task flow graph (ETFG) of the considered application. |
| $f_{\text{node}}$ | Function that transforms a node $N_i \in G$ into a composite node $N'_i \in G'$. |
| $f_{\text{arc}}$ | Function that transforms an arc $A_{i \to j} \in G$ into a composite arc $A'_{i \to j} \in G'$. |
| $\mathcal{N}'$ | Set of composite nodes in the ETFG. |
| $\mathcal{A}'$ | Set of composite arcs in the ETFG. |
| $N'_i$ | Composite node in the ETFG ($N'_i \in \mathcal{N}'$) comprising a set of candidate nodes $N_{ik}$. |
| $N_{ik}$ | Candidate node in $N'_i$ corresponding to the allocation of task $i$ on device $k$. |
| $A'_{i \to j}$ | Composite arc in the ETFG ($A'_{i \to j} \in \mathcal{A}'$) comprising a set of arcs $A_{ik \to jl}$. |
| $A_{ik \to jl}$ | Arc in $A'_{i \to j}$ corresponding to the data flow between candidate nodes $N_{ik}$ and $N_{jl}$. |
| $L_{ik}$ | Computational latency (execution time) of task $i$ on device $k$. |
| $P_{ik}$ | Computational power consumption required to execute task $i$ on device $k$. |
| $E_{ik}$ | Computational energy consumption required to execute task $i$ on device $k$. |
| $M_i$ | Main memory requirement of task $i$. |
| $S_i$ | Storage requirement of task $i$. |
| $D_i$ | Output data size of task $i$. |
| $NC_i$ | Number of child tasks of task $i$. |
| $CL_{ik \to jl}$ | Communication latency of arc $A_{ik \to jl}$. |
| $CE_{ik \to jl}$ | Communication energy consumption of arc $A_{ik \to jl}$. |
| $\delta^m_{ik \to jl}$ | Binary parameter denoting whether arc $A_{ik \to jl}$ involves indirect communication through device $m$. |
| $W_{kl}$ | Bandwidth of communication channel between devices $k$ and $l$. |
| $\tau_{kl}$ | Energy required to transmit a unit of data from device $k$ to device $l$. |
| $\rho_{kl}$ | Energy required to receive a unit of data at device $l$ from device $k$. |
| $x_{ik}$ | Binary decision variable corresponding to candidate node $N_{ik}$. |
| $x_{ik \to jl}$ | Binary decision variable corresponding to arc $A_{ik \to jl}$. |
| $M_k^{\text{bgt}}$ | Memory budget corresponding to device $k$. |
| $S_k^{\text{bgt}}$ | Storage budget corresponding to device $k$. |
| $E_k^{\text{bgt}}$ | Energy budget corresponding to device $k$. |
| $L_{\text{thr}}$ | Overall latency threshold. |
| $\phi_k$ | Performance ratio of device $k$ for a specific device benchmark with respect to reference device ê.* |
| $\phi_k^{\text{avg}}$ | Average performance ratio of device $k$ across all device benchmarks with respect to reference device ê.* |
| $P_k^{\text{idle}}$ | Idle power consumption of device $k$.* |
| $P_k^{\text{max}}$ | Maximum power consumption of device $k$.* |
| $\alpha$ | Adjustment factor of computational power consumption.* |

*Used in the transformation of random TFGs into ETFGs.

*Corresponding author.
  *Email address:* kouloumpris.andreas@ucy.ac.cy (Andreas Kouloumpris)

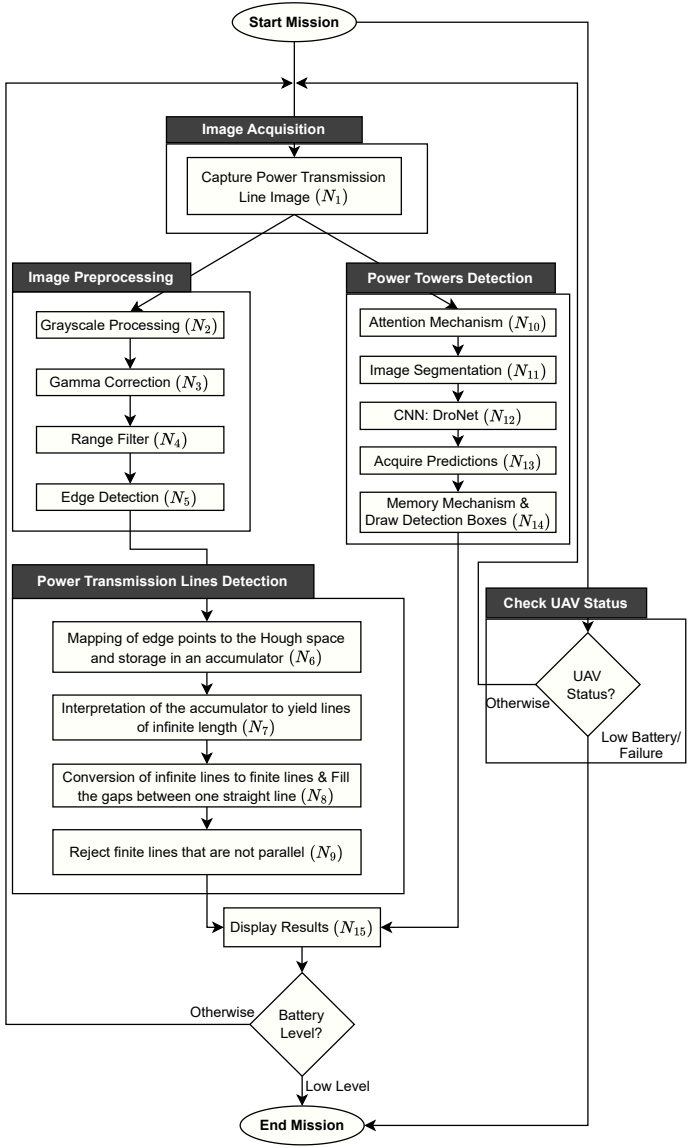 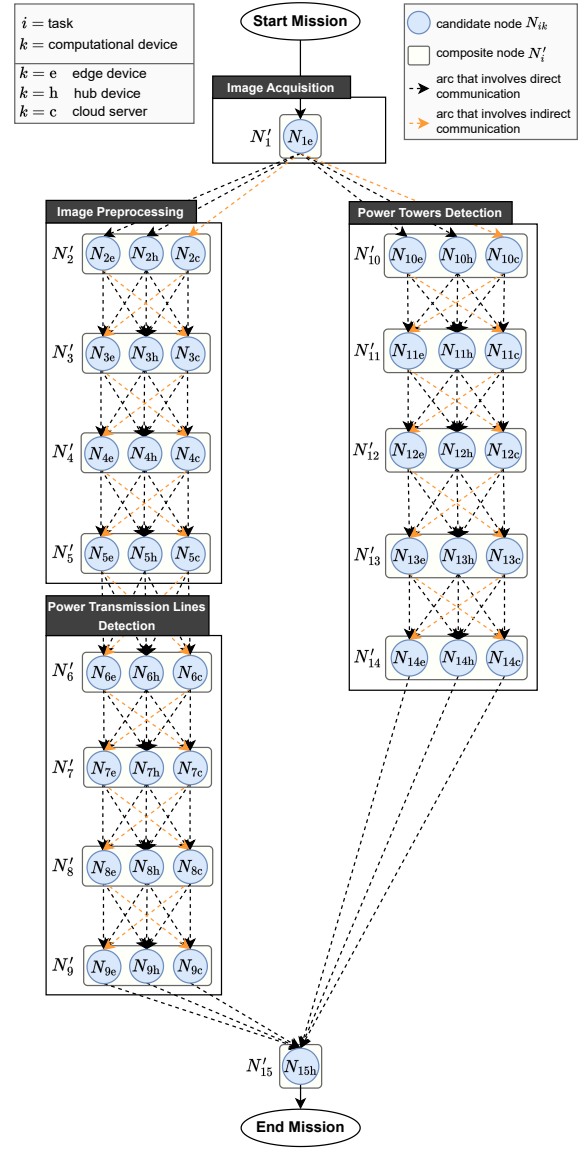

(a) Flowchart of the application encapsulating its TFG

(b) Corresponding ETFG

Figure A.1: Real-world use-case application: (a) its flowchart that encapsulates its TFG, and (b) its corresponding ETFG.



\end{document}